\documentclass[trackchanges,twocolumn]{aastex701}

\newcommand{\src}{1ES~1927+654}
\defcitealias{Masterson2025}{M25}
\usepackage{caption}
\usepackage[shortlabels]{enumitem}

\begin{document}

\title{Persistence of the Millihertz X-ray Quasi-Periodic Oscillation in the Active Galactic Nucleus 1ES~1927+654}
\shorttitle{The Millihertz X-ray QPO in 1ES~1927+654}

\author[0000-0003-4127-0739]{Megan Masterson}
\affiliation{MIT Kavli Institute for Astrophysics and Space Research,
Massachusetts Institute of Technology, 
Cambridge, MA 02139, USA}
\email[show]{mmasters@mit.edu}
\correspondingauthor{Megan Masterson}

\author[0000-0003-0172-0854]{Erin Kara}
\affiliation{MIT Kavli Institute for Astrophysics and Space Research, Massachusetts Institute of Technology, Cambridge, MA 02139, USA}
\email[]{ekara@mit.edu}

\author[0000-0003-2658-6559]{William N. Alston}
\affiliation{Centre for Astrophysics Research, Department of Physics, Astronomy and Mathematics, University of Hertfordshire, College Lane, Hatfield, AL10 9AB, UK}
\email[]{w.alston@herts.ac.uk}

\author[0000-0003-4054-7978]{Riccardo Arcodia}
\affiliation{Black Hole Initiative at Harvard University, 20 Garden Street, Cambridge, MA 02138, USA}
\email[]{rarcodia@fas.harvard.edu}

\author[0000-0003-0936-8488]{Mitch Begelman}
\affiliation{JILA, University of Colorado and National Institute of Standards and Technology, 440 UCB, Boulder, CO 80309-0440, USA}
\affiliation{Department of Astrophysical and Planetary Sciences, University of Colorado, 391 UCB, Boulder, CO 80309-0391, USA}
\email[]{mitch@jila.colorado.edu}

\author[0000-0002-0568-6000]{Joheen Chakraborty}
\affiliation{MIT Kavli Institute for Astrophysics and Space Research, Massachusetts Institute of Technology, Cambridge, MA 02139, USA}
\email[]{joheen@mit.edu}

\author[0000-0002-9378-4072]{Andrew C. Fabian}
\affiliation{Institute of Astronomy, University of Cambridge, Madingley Road, Cambridge CB3 0HA, UK}
\email[]{acf@ast.cam.ac.uk}

\author[0000-0002-1329-658X]{Margherita Giustini}
\affiliation{Centro de Astrobiolog\'{i}a (CAB), CSIC-INTA, Camino Bajo del Castillo s/n, Villanueva de la Cañada, 28692 Madrid, Spain}
\email[]{mgiustini@cab.inta-csic.es}

\author[0000-0002-5311-9078]{Adam Ingram}
\affiliation{School of Mathematics, Statistics, and Physics, Newcastle University, Newcastle upon Tyne NE1 7RU, UK}
\email[]{adam.ingram@newcastle.ac.uk}

\author[0000-0003-4511-8427]{Peter Kosec}
\affiliation{Center for Astrophysics $\mid$ Harvard \& Smithsonian, Cambridge, MA 02138, USA}
\email[]{peter.kosec@cfa.harvard.edu}

\author[0000-0003-2714-0487]{Sibasish Laha} 
\affiliation{Astrophysics Science Division, NASA Goddard Space Flight Center, Greenbelt, MD 20771, USA.}
\affiliation{Center for Space Science and Technology, University of Maryland Baltimore County, 1000 Hilltop Circle, Baltimore, MD 21250, USA.}
\affiliation{Center for Research and Exploration in Space Science and Technology, NASA/GSFC, Greenbelt, Maryland 20771, USA}
\email[]{sibasish.laha@nasa.gov}

\author[0000-0003-0707-4531]{Giovanni Miniutti}
\affiliation{Centro de Astrobiolog\'{i}a (CAB), CSIC-INTA, Camino Bajo del Castillo s/n, Villanueva de la Cañada, 28692 Madrid, Spain}
\email[]{gminiutti@cab.inta-csic.es}

\author[0009-0001-9034-6261]{Christos Panagiotou}
\affiliation{MIT Kavli Institute for Astrophysics and Space Research, Massachusetts Institute of Technology, Cambridge, MA 02139, USA}
\email[]{cpanag@mit.edu}

\author[0000-0003-2532-7379]{Ciro Pinto}
\affiliation{INAF - IASF Palermo, Via U. La Malfa 153, I-90146 Palermo, Italy}
\email[]{ciro.pinto@inaf.it}

\author[0000-0001-5231-2645]{Claudio Ricci}
\affiliation{Department of Astronomy, University of Geneva, ch. d’Ecogia 16, 1290, Versoix, Switzerland}
\affiliation{Instituto de Estudios Astrof\'{i}sicos, Facultad de Ingenier\'{i}a y Ciencias, Universidad Diego Portales, Av. Ej\'{e}rcito Libertador 441, Santiago, Chile}
\email[]{claudio.ricci@unige.ch}
	
\author[0000-0002-9163-8653]{Dev R. Sadaula} 
\affiliation{Astrophysics Science Division, NASA Goddard Space Flight Center, Greenbelt, MD 20771, USA.}
\affiliation{Center for Space Science and Technology, University of Maryland Baltimore County, 1000 Hilltop Circle, Baltimore, MD 21250, USA.}
\affiliation{Center for Research and Exploration in Space Science and Technology, NASA/GSFC, Greenbelt, Maryland 20771, USA}
\email[]{dev.r.sadaula@nasa.gov}

\author[0000-0003-4727-2209]{Onic I. Shuvo}
\affiliation{Department of Physics, University of Maryland Baltimore County, 1000 Hilltop Circle Baltimore, MD 21250, USA}
\email[]{oishuvo@umbc.edu}

\author[0000-0002-3683-7297]{Benny Trakhtenbrot}
\affiliation{School of Physics and Astronomy, Tel Aviv University, Tel Aviv 69978, Israel}
\email[]{bennyt@tauex.tau.ac.il}

\begin{abstract}

1ES~1927+654 is an extreme active galactic nucleus (AGN) that has defied our canonical expectations for how AGN appear across the electromagnetic spectrum and how they vary on short timescales. In 2022, this source began showing a X-ray quasi-periodic oscillation (QPO) at mHz frequencies, along with a newly launched radio jet. Unlike the handful of other known AGN QPOs, the QPO in 1ES~1927+654 showed a significant frequency evolution, spanning from 0.9-2.4 mHz from 2022-2024. In this work, we present the last 1.5 years of monitoring with XMM-Newton (250 ks) up to January 2026, which reveals that the QPO persists but has plateaued at a constant frequency of approximately 2.5 mHz. We perform detailed spectral-timing analyses on this exquisite dataset, consisting of over 900 QPO cycles, more than any AGN QPO to date. Our main findings are: (1) the stacked XMM-Newton power spectra shows no significant second harmonic, (2) a soft (reverberation-like) lag is observed at all frequencies and remains remarkably stable even as the QPO frequency evolved from 2022-2024, and (3) extreme X-ray jumps on the QPO period (up to $\sim$80\% baseline flux) persist to present day with a remarkably stable dip-rise-fall pattern. Finally, we also detect the first AGN QPO in NuSTAR observations, which is present from 2023 to 2026 at frequencies consistent with the XMM-Newton detections. While we explore models for eclipses and coupled disk-corona behavior to simultaneously explain the lags, dips, and QPO, these new observations strain such models. 

\end{abstract}

\keywords{\uat{Active galactic nuclei}{16} --- \uat{High Energy astrophysics}{739} --- \uat{Supermassive black holes}{1663} --- \uat{X-ray active galactic nuclei}{2035}}


\section{Introduction} \label{sec:intro}

Quasi-periodic oscillations (QPOs) are coherent, regular variability on top of the stochastic variability seen in accreting black holes. Despite being abundantly detected in black hole X-ray binaries \citep[BHXRBs; see][for a recent review]{Ingram2019} throughout the galaxy, these features are notoriously difficult to detect in accreting supermassive black holes \citep[SMBHs; see, e.g.,][]{Vaughan2005a,Gonzalez-Martin2012}. Historically, only a handful of SMBHs show robust QPO detections \citep[e.g.,][]{Gierlinski2008,Alston2014}. Assuming that accretion is scale invariant, QPO frequencies should scale inversely with the mass of the black hole; hence, we expect that the only QPOs that are easily detected in single, continuous X-ray observations of SMBHs ($\sim 100$~ks) are scaled-up versions of the rare high-frequency QPOs (HFQPOs; $f_\mathrm{QPO} \sim 50-500$~Hz) in BHXRBs. Due to their relatively constant frequency that corresponds to motion near the innermost stable circular orbit (ISCO), it has long been believed that HFQPOs encode information about the black hole mass and spin \citep[e.g.,][]{Morgan1997,Remillard1999,Remillard2002,Abramowicz2001,Motta2014b,Motta2014a}. However, using these features as direct probes of these fundamental quantities requires a proper understanding of the mechanism producing them. 


\src\, is one of only a handful of SMBHs that shows a robust, persistent QPO \citep[][hereafter \citetalias{Masterson2025}]{Masterson2025}. This source has a storied multi-wavelength history that is pertinent to understanding the onset of this QPO. Starting in 2018, \src\, went through a major outburst marked by an increase in optical flux of $\sim$4 mag and the formation of prominent broad Balmer emission lines on timescales of months \citep{Trakhtenbrot2019}. Follow-up X-ray observations taken a few months after the peak of the optical flare revealed a soft, thermal X-ray spectrum \citep{Ricci2020}, with no signatures of the standard hard X-ray power law seen previously in \src\, \citep{Gallo2013}. This indicated that the canonical X-ray corona that is ubiquitous in active galactic nuclei (AGN) had been destroyed, which has been suggested to be related to a tidal disruption event (TDE) occurring in an existing AGN \citep{Ricci2020,Ricci2021,Cao2023} or an inward-propagating magnetic flux inversion \citep{Scepi2021,Laha2022}. Over the subsequent 3 years, the intrinsic 0.3-10 keV X-ray luminosity of this source varied by four orders of magnitude, including order of magnitude variability on timescales of hours, as the corona was reignited \citep{Ricci2021,Masterson2022}. \src\, returned to its pre-outburst X-ray flux and spectral state by mid-2021 \citep{Masterson2022,Laha2022}, seemingly marking the end of the extreme outburst. However, beginning in 2022, \cite{Ghosh2023} noted that the soft X-ray flux started to increase again in this source, notably with no optical or UV counterpart. The first detection of the mHz-frequency X-ray QPO occurred shortly thereafter, in mid-2022 with XMM-Newton \citepalias{Masterson2025}. Less than a year later, the radio flux also showed a dramatic increase, and VLBI observations revealed an extended, outflowing lobes that are thought to arise from a newly launched jet \citep{Meyer2025,Laha2025}. These observations imply that the outburst of \src\, is far from over and hint at a potential connection between the newly formed corona, the X-ray QPO, and the newly launched radio jet. 

One of the most unique aspects of the mHz-frequency X-ray QPO in \src\, is that the frequency showed significant variability \citepalias{Masterson2025}. In 2022, the QPO was detected at $\approx 0.9$ mHz, but within 6 months the frequency rose to 1.67 mHz. Within the next year, the frequency continued to increase, but appeared to be stalling around 2.4 mHz (i.e., $\dot{f} > 0$, $\ddot{f} < 0$, where $f$ is the QPO frequency). Such evolution had never been seen before. The best-studied SMBH QPO in RE J1034+396 has shown just 10\% fluctuations in the QPO frequency over two decades of observations \citep{Gierlinski2008,Alston2014,Xia2024}. Likewise, an X-ray QPO detected in the TDE ASASSN-14li remained extremely stable in frequency over 400 days and more than an order of magnitude change to the X-ray flux \citep{Pasham2019}. The constant frequency in both of these sources is roughly consistent with the expectation that the observed AGN QPOs are scaled-up version of HFQPOs in BHXRBs, as these are believed to be fundamental and related to the mass and spin of the black hole. Thus, the QPO in \src\, challenges current models for SMBH QPOs and opens a new window into our understanding of scale-invariant accretion processes. 

In this work, we present continued monitoring of \src\, with XMM-Newton, focusing on the evolution of the QPO from mid-2024 to early 2026 and the implications of these new observations on our understanding of the driving mechanism for the QPO. In Section \ref{sec:obs}, we present details for the 18 new XMM-Newton observations over this period, as well as 4 archival NuSTAR observations from 2023 that also show the QPO. The resulting QPO evolution is presented in Section \ref{sec:results}, together with X-ray time lags that we use to diagnose the origin of the variability. We discuss the implications of these findings with respect to QPO models and the broader picture of this source in Section \ref{sec:discussion}, and finally, we summarize our findings in Section \ref{sec:conclusion}. Throughout this work, we assume a standard flat, $\Lambda$CDM cosmology with $\Omega_M = 0.7$ and $H_0 = 70$ km s$^{-1}$ Mpc$^{-1}$. All quoted uncertainties represent 1$\sigma$ confidence (encompassing 68\% confidence intervals) unless otherwise noted. 


\section{Observations and Data Reduction} \label{sec:obs}

Table \ref{tab:obs} shows the observation details for the new XMM-Newton and NuSTAR data that are presented in this manuscript. We also include in our analysis the epochs of observations from \citetalias{Masterson2025} with a strong QPO detection (i.e., ObsIDs 0915390701, 0931791401, and 0932392001), the details of which are provided in Extended Data Tables 1 and 2 of that work.

\begin{deluxetable*}{ccccccccc}
    \centering
    \caption{QPO Parameters and New Observation Details}
    \tablehead{\colhead{Observatory} & \colhead{ObsID} & \colhead{Date} & \colhead{Exposure\tablenotemark{$a$}} & \colhead{$f_\mathrm{QPO}$} & \colhead{QPO Significance\tablenotemark{$b$}} & \colhead{$Q$\tablenotemark{$c$}} & \colhead{RMS$_\mathrm{QPO}$\tablenotemark{$d$}} & \colhead{N$_\mathrm{cycles}$\tablenotemark{$e$}} \\ &  &  & \colhead{(ks)} & \colhead{(mHz)} & \colhead{($\sigma$)} & \colhead{} & \colhead{} & \colhead{}}
    \startdata
    XMM-Newton & 0953010401 & 2024-07-19 & 22.1 & $2.44^{+0.11}_{-0.09}$ & 4.9 & $6^{+4}_{-1}$ & $0.13^{+0.03}_{-0.02}$ & 54 \\
& 0953010501 & 2024-07-27 & 24.3 & $2.44^{+0.09}_{-0.07}$ & 4.2 & $6^{+5}_{-2}$ & $0.10^{+0.02}_{-0.02}$ & 59 \\
& 0953010901 & 2024-10-13 & 8.8 & $2.35^{+0.17}_{-0.11}$ & 2.6 & $6^{+11}_{-2}$ & $0.14^{+0.05}_{-0.04}$ & 21 \\
& 0953010601 & 2024-10-21 & 20.1 & $2.43^{+0.06}_{-0.05}$ & 6.9 & $9^{+9}_{-3}$ & $0.15^{+0.03}_{-0.02}$ & 49 \\
& 0953010801 & 2025-01-19 & 22.5 & $2.52^{+0.04}_{-0.04}$ & 5.6 & $14^{+17}_{-6}$ & $0.11^{+0.03}_{-0.02}$ & 57 \\
& 0953010701 & 2025-01-25 & 18.6 & $2.53^{+0.05}_{-0.05}$ & 4.3 & $15^{+31}_{-8}$ & $0.09^{+0.03}_{-0.02}$ & 47 \\
& 0970190101 & 2025-04-30 & 17.9 & $2.54^{+0.07}_{-0.07}$ & 6.5 & $6^{+4}_{-2}$ & $0.13^{+0.02}_{-0.02}$ & 45 \\
& 0970190301 & 2025-05-03 & 20.1 & $2.51^{+0.07}_{-0.06}$ & 4.5 & $10^{+14}_{-5}$ & $0.12^{+0.03}_{-0.02}$ & 50 \\
& 0970190201 & 2025-05-05 & 16.5 & $2.54^{+0.06}_{-0.08}$ & 5.6 & $8^{+12}_{-3}$ & $0.14^{+0.03}_{-0.03}$ & 42 \\
& 0970190401 & 2025-07-07 & 16.6 & $2.58^{+0.07}_{-0.06}$ & 3.6 & $11^{+15}_{-5}$ & $0.11^{+0.03}_{-0.02}$ & 43 \\
& 0970190501 & 2025-07-09 & 15.2 & $2.76^{+0.30}_{-0.20}$ & 3.7 & $6^{+34}_{-1}$ & $0.14^{+0.05}_{-0.03}$ & 42 \\
& 0970190601 & 2025-08-02 & 13.5 & $2.40^{+0.08}_{-0.10}$ & 2.7 & $10^{+31}_{-5}$ & $0.10^{+0.04}_{-0.03}$ & 32 \\
& 0970190701 & 2025-10-21 & 16.2 & $2.70^{+0.11}_{-0.11}$ & 4.3 & $6^{+4}_{-1}$ & $0.17^{+0.04}_{-0.03}$ & 44 \\
& 0970190801\tablenotemark{$f$} & 2025-10-25 & 8.1 & --- & --- & --- & --- & --- \\
& 0970190901 & 2025-10-29 & 16.1 & $2.46^{+0.08}_{-0.13}$ & 2.8 & $7^{+21}_{-3}$ & $0.11^{+0.04}_{-0.03}$ & 40 \\
& 0970191001 & 2026-01-07 & 21.1 & $2.48^{+0.07}_{-0.06}$ & 5.6 & $8^{+13}_{-3}$ & $0.13^{+0.04}_{-0.02}$ & 52 \\
& 0970191101\tablenotemark{$g$} & 2026-01-09 & --- & --- & --- & --- & --- & --- \\
& 0970191201 & 2026-01-11 & 15.3 & $2.55^{+2.45}_{-0.55}$ & 1.1 & $15^{+339}_{-9}$ & $0.07^{+0.06}_{-0.04}$ & 39 \\
    \hline
    NuSTAR & 80902632002 & 2023-05-25 & 37.1 & $2.14^{+0.08}_{-0.06}$ & 2.9 & $5^{+7}_{-2}$ & $0.16^{+0.03}_{-0.03}$ & 319 \\
& 80902632004 & 2023-06-24 & 41.1 & & & & & \\
& 80902632006 & 2023-08-03 & 39.8 & & & & & \\
& 80902632008 & 2023-09-02 & 31.0 & & & & & \\
\cline{2-9}
& 61002008002 & 2024-06-10 & 38.8 & $2.40^{+0.13}_{-0.07}$ & 2.6 & $9^{+23}_{-4}$ & $0.14^{+0.05}_{-0.03}$ & 393 \\
& 61002008004 & 2024-07-29 & 40.9 & & & & & \\
& 61002008006 & 2024-09-19 & 41.1 & & & & & \\
& 61002008008 & 2024-12-16 & 43.0 & & & & & \\
\cline{2-9}
& 61002008010 & 2025-01-31 & 50.5 & $2.59^{+0.04}_{-0.08}$ & 3.3 & $15^{+76}_{-8}$ & $0.12^{+0.10}_{-0.03}$ & 473 \\
& 61002008012 & 2025-03-13 & 43.4 & & & & & \\
& 61102017002 & 2025-06-02 & 47.8 & & & & & \\
& 61102017004 & 2025-08-04 & 41.2 & & & & & \\
\cline{2-9}
& 61102017006 & 2025-10-15 & 44.2 & $2.42^{+0.10}_{-0.04}$ & 3.2 & $13^{+55}_{-8}$ & $0.16^{+0.15}_{-0.04}$ & 326 \\
& 61102017008 & 2025-12-21 & 45.6 & & & & & \\
& 61102017010 & 2026-02-15 & 44.7 & & & & & \\
    \enddata
    \label{tab:obs}
    \tablenotetext{a}{For XMM-Newton data, this is the cleaned exposure time, after the removal of periods of strong background flaring.}
    \tablenotetext{b}{This is the significance of the QPO feature (an additional Lorentzian) using the power-law broadband noise model, as estimated with $\Delta$AIC.}
    \tablenotetext{c}{This is the quality factor of the QPO feature, defined as $Q = f_\mathrm{QPO}/\mathrm{FWHM}_\mathrm{QPO}$.}
    \tablenotetext{c}{The fractional RMS denotes the strength of the QPO feature, which is measured based on the normalization of the additional Lorentzian feature for the QPO.}
    \tablenotetext{e}{This denotes the number of QPO cycles in this observation, as found by multiplying the QPO frequency by the exposure duration.}
    \tablenotetext{f}{Bad background flaring left only 8.1 ks of data for this observation, and no significant improvement to the PSD is made by including an additional Lorentzian (i.e., the QPO). Hence, we do not include this observation in our analysis.}
    \tablenotetext{g}{This observation was plagued by extremely bad background flaring, leaving less than 5~ks of usable continuous data. Hence, we do not include this observation in our analysis.}
\end{deluxetable*}

\subsection{XMM-Newton}

XMM-Newton has extensively observed \src, including a pre-outburst observation in 2011 and 33 subsequent observations since the beginning of its 2018 outburst. In this work, we focus on the X-ray variability in the 18 newest observations, which were taken between 2024 July and 2026 January. We reduced the EPIC-pn data using the XMM-Newton Science Analysis System (version 20.0.0) with the latest calibration files and followed standard data reduction procedures, including running \texttt{epproc} to produce calibrated event files, removing periods of significant background flaring, and creating RMF and ARF files with \texttt{rmfgen} and \texttt{arfgen}, respectively. To facilitate Fourier timing analysis, we aimed to maximize the amount of continuous data we could obtain by cutting out the last $\sim$10-30\% of each observation, which contains the most significant background flaring. We extracted source counts from a circular region with a radius of 35\arcsec\, and background counts from a circular region located away from the source with a radius of 40\arcsec. There is only mild evidence for pile-up effects, and hence we use a circular extraction region to improve the signal-to-noise in the 2-10 keV band for timing purposes. For all spectral analysis, we excise the inner 10\arcsec\, to mitigate even mild pile-up effects. Finally, we construct background-subtracted light curves with the \texttt{epiclccorr} tool, using 20s time bins to capture the rapid variability in this system.

\subsection{NuSTAR}

In this work, we investigate the variability in the NuSTAR observations taken between 2023 and 2026 (PI: S. Laha). The details of each observation are given in Table \ref{tab:obs}. We reduced these observations using NuSTAR Data Analysis Software (NuSTARDAS; version 2.1.2 in HEASoft version 6.33.2) with calibration files from NuSTAR CALDB v20220706. We followed standard data reduction procedures for NuSTAR data, which include processing the data with \texttt{nupipeline} and extracting light curves for both the FPMA and FPMB modules with \texttt{nuproducts}. When creating light curves, we used a circular region centered on the source with a radius of 90\arcsec and a circular background region located away from the source with a radius of 120\arcsec.

\section{Results} \label{sec:results}

\subsection{QPO Detection and Significance}

To assess the evolution of the QPO, we constructed power spectral densities (PSDs) for each of the observations with [rms/mean]$^2$ normalization \citep{Vaughan2003} using the \texttt{pyLag}\footnote{\url{https://github.com/wilkinsdr/pylag}} spectral-timing package \citep{Wilkins2019}. The full hard X-ray (2-10 keV) light curves and their respective PSDs are shown in Figure \ref{fig:all_lc_psd} of the Appendix. In this work, we follow the same $\Delta$AIC methodology as in \citetalias{Masterson2025}. In brief, for all XMM-Newton data we work with the unbinned PSDs, performing maximum likelihood estimation with the Whittle likelihood, given by
\begin{equation} \label{eqn:whit_like}
    \ln\mathcal{L} = -\sum_i \left[\frac{I_i}{S_i} - \ln(S_i)\right],
\end{equation}
where $i$ denotes a sum over all frequencies, $I_i$ is the measured power at a frequency $f_i$, and $S_i$ is the model power at a frequency $f_i$. This likelihood applies for variable that follow a $\chi^2_2$ distribution, as discussed in detail in \cite{Vaughan2005b,Vaughan2010}. We compare the resulting likelihoods using the sample-corrected Akaike Information Criterion \citep[AIC;][]{Akaike1974}, which is given by 
\begin{equation}
    \mathrm{AIC}_c = 2k - 2C_L - 2\ln\mathcal{L} + \frac{2 k (k+1)}{N-k-1}
\end{equation}
where $C_L$ is the (unknown) likelihood of the true model, $k$ is the number of model free parameters, $N$ is the number of frequency points, and $\ln \mathcal{L}$ is the maximum likelihood defined by Equation \ref{eqn:whit_like}. The statistical improvement of one model over another (e.g., the addition of a QPO on the broadband noise) can then be computed with $p_\mathrm{AIC} = e^{-\Delta\mathrm{AIC}/2}$. For a detailed description of this methodology, we refer the reader to \citetalias{Masterson2025}.

\begin{figure*}[t!]
    \centering
    \includegraphics[width=\linewidth]{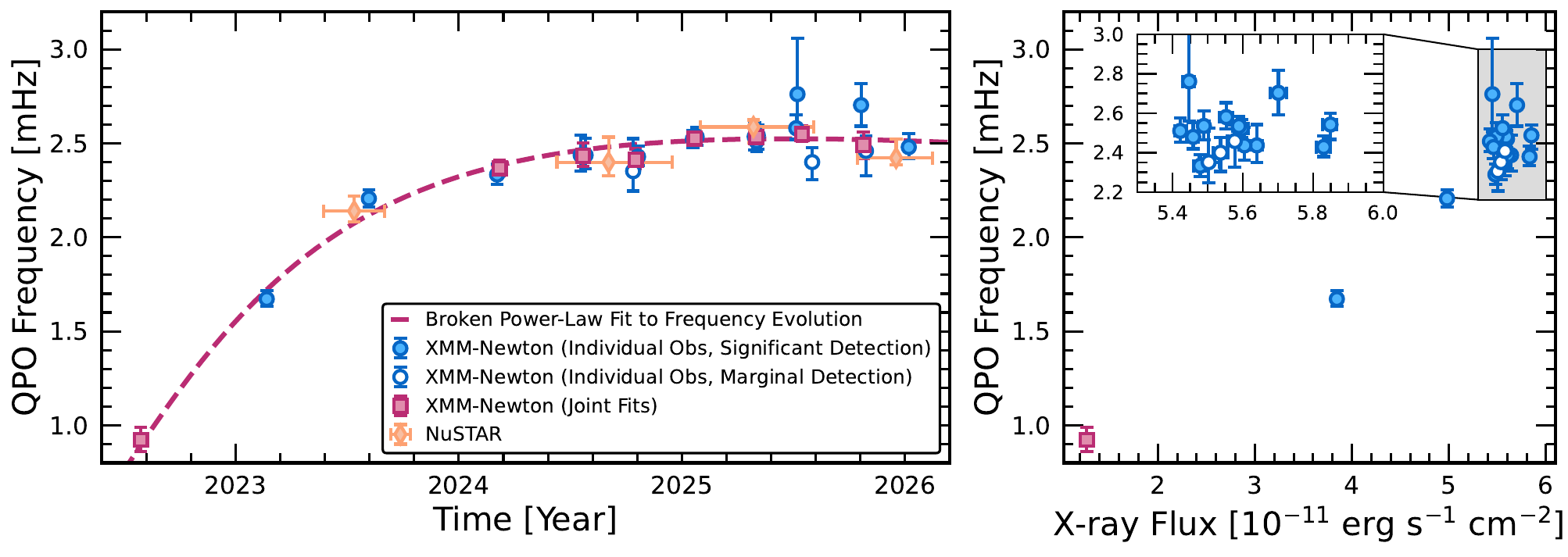}
    \caption{Evolution of the QPO frequency. \textit{Left:} Evolution of the QPO frequency with time. The blue circles show XMM-Newton observations in the 2-10 keV band, with filled circles denoting significant detections ($> 3\sigma$) and open circles denoting  marginal detections ($2-3\sigma$), as estimated using $\Delta$AIC. The pink squares show joint fits to multiple XMM-Newton observations in a given epoch, which all show significant detections with improved measurement of the QPO frequency. The orange diamonds show the NuSTAR detections in the 3-10 keV band. The QPO is persistent, with only mild frequency evolution since early 2024 ($\approx 0.1$ mHz). The pink dashed line shows the best-fit bending power-law model to the frequency evolution with the XMM-Newton data. \textit{Right:} Evolution of the QPO frequency with 0.3-10 keV X-ray flux. The observations are still broadly consistent with the QPO frequency scaling monotonically with the total X-ray flux, although, as the zoom-in shows, there is no strong frequency dependence in the last $\sim$1.5 years of data.}
    \label{fig:qpo_freq_evol}
\end{figure*}

For all XMM-Newton observations, the significance of the feature is robust against choices for the shape of the broadband noise. We tested both a simple power-law and a broken power-law model with the low frequency slope fixed at $\alpha = 0$, finding that the two provide comparable results. All reported fit parameters from the individual and joint fits (e.g., Figure \ref{fig:qpo_freq_evol} and Table \ref{tab:obs}) are the results for a power-law broadband noise model, as the statistical significance for a break in the individual 2-10 keV PSDs is low ($\ll 3\sigma$ in most observations). We report significant detections of the QPO as those with a $\geq 3\sigma$ improvement upon the power-law broadband model with the addition of a QPO, while we define ``marginal" detections as those with statistical significance of the QPO between $2-3\sigma$. In total, we report 12 new significant detections, 2 marginal detections, and 1 non-detection in 15 observations with $> 10$~ks of continuous data.

\begin{figure}
    \centering
    \includegraphics[width=\linewidth]{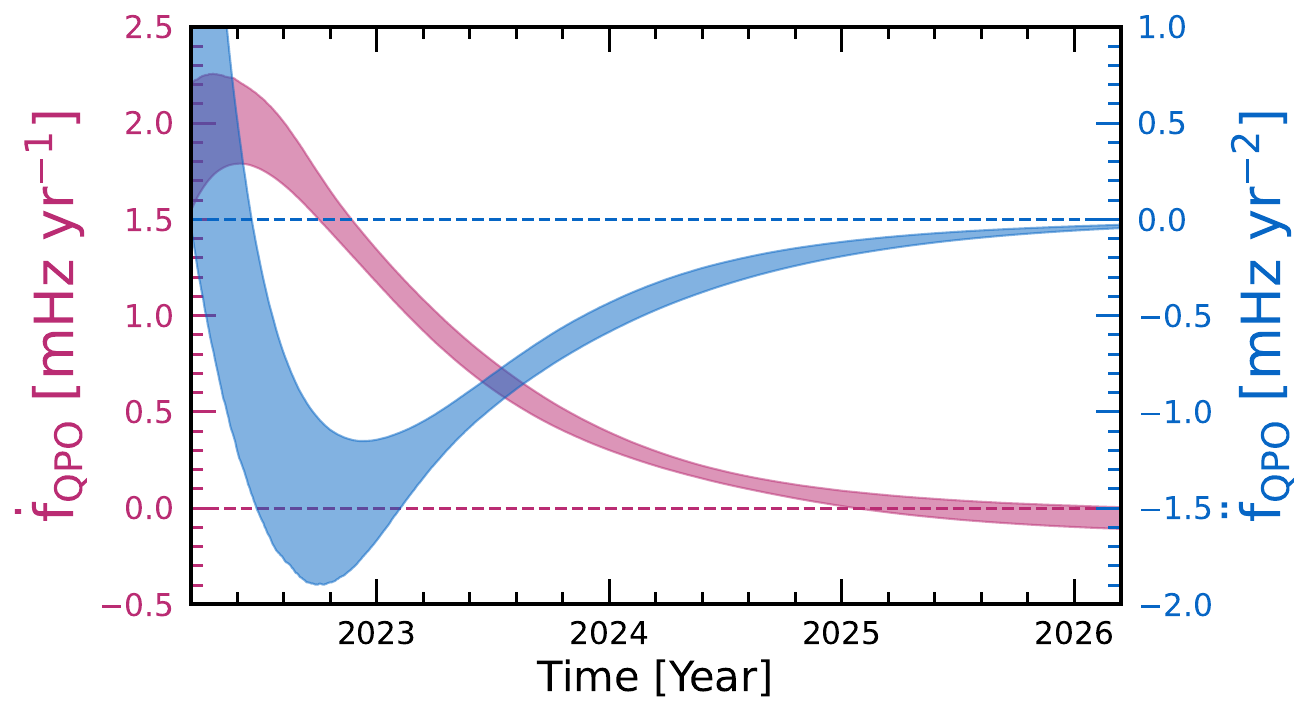}
    \caption{Derivatives of the QPO frequency with respect to time (i.e., $\dot{f}_\mathrm{QPO}$ and $\ddot{f}_\mathrm{QPO}$) based on the best-fit bending power-law model for the QPO frequency evolution. The shaded pink (blue) region shows the 1$\sigma$ confidence interval for the $\dot{f}_\mathrm{QPO}$ ($\ddot{f}_\mathrm{QPO}$) evolution based on the covariance matrix for the fit to the model.}
    \label{fig:qpo_freq_derivs}
\end{figure}

Figure \ref{fig:qpo_freq_evol} shows the evolution of the QPO frequency with time (left) and the X-ray flux (right). The QPO frequency continues to follow a significant trend with the total X-ray flux, albeit with some scatter in the most recent observations. Additionally, the QPO frequency is more stable in the last 1.5 years than prior to 2024. To assess the temporal evolution of the QPO frequency, we fit the data with a bending power-law, given by 
\begin{equation}
    f_\mathrm{QPO}(t) = A \left(\frac{t}{t_b}\right)^{\beta_1} \left[1 + \left(\frac{t}{t_b}\right)^{-\beta_2 + \beta_1}\right]^{-1},
\end{equation}
where $A$ is a constant, $t_b$ is the bending time, $\beta_1$ is the early-time slope, and $\beta_2$ is the late-time slope. The best-fit values are $A = 4.1 \pm 0.8$ mHz, $t_b = 2023.5 \pm 0.6$, $\beta_1 = 1.4 \pm 0.3$, and $\beta_2 = -0.3 \pm 0.2$. Figure \ref{fig:qpo_freq_derivs} shows the resulting $\dot{f}_\mathrm{QPO}$ and $\ddot{f}_\mathrm{QPO}$ based on the best-fit bending power-law model, with the bands representing the 1$\sigma$ uncertainty regions based on 1000 random samples drawn from the best-fit covariance matrix. Given all parameter uncertainties, the QPO frequency is currently consistent with $\dot{f}_\mathrm{QPO} = 0$. 

\begin{figure*}
    \centering
    \includegraphics[width=\linewidth]{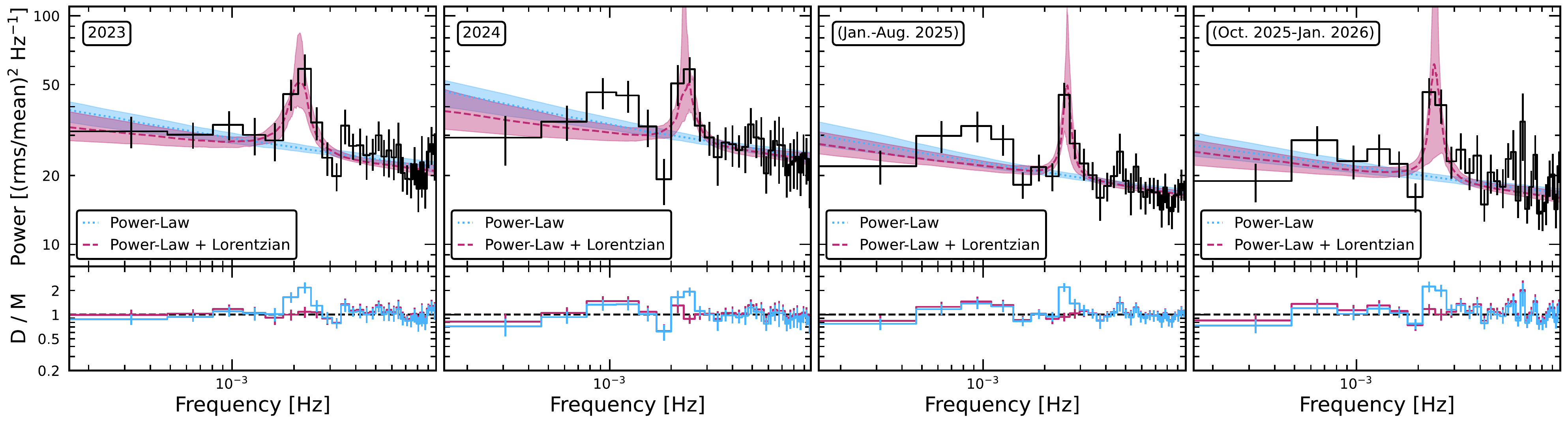}
    \caption{Existence of a QPO in the NuSTAR data from 2023-2026. The top panel of each plot shows the stacked NuSTAR PSD in the 3-10~keV band with 50s time bins. The error bars correspond to the standard error on the mean. The blue dotted line shows the broadband model, assuming only a power-law and constant. The pink dashed line shows the addition of a Lorentzian for the QPO on top of this broadband model. For both models, the shaded region shows the 1$\sigma$ uncertainty region from MCMC chains. The bottom panel shows the data divided by model for both the broadband only (pink) and broadband plus QPO model (blue). A clear QPO is evident in all stacks at comparable frequencies to what is measured in individual XMM-Newton observations.}
    \label{fig:nustar_psd_all}
\end{figure*}

The low Earth orbit of NuSTAR impedes our ability to apply the same simple timing techniques as with XMM-Newton data. This orbit gives rise to regular gaps in the light curves every 3-3.5~ks, corresponding to times in which the source is behind the Earth. Despite this difficulty, we can still detect the QPO in \src, as its frequency is about one order of magnitude higher than the minimum frequency probed by a continuous $\sim$3~ks segment. The relatively low count rate of NuSTAR observations ($\approx 0.1$ cps for \src\, in the 3-10 keV band) means that we must combine PSDs computed from numerous continuous segments of data to confidently detect this feature. Thus, to increase the signal-to-noise, we add the signal from the FPMA and FPMB detectors, create one PSD from each continuous segment longer than 3~ks, and finally combine the PSDs from all segments across many observation. We note that the NuSTAR dead time ($\sim2.5$ ms) is significantly shorter than the timescales of interest in \src, thereby enabling this rather simple analysis compared to the cross-spectral analyses required in BHXRBs when the dead time is comparable to the timescales of interest \citep{Bachetti2015}. After creating the PSDs, we stack them on roughly 6 month cadence, finding that roughly 150~ks of total exposure time is needed to detect the QPO at 3$\sigma$ significance. Figure \ref{fig:nustar_psd_all} shows the resulting PSDs from the four stacks with 50s time bins. A clear peak in the PSD is evident in each stack at a similar frequency as seen by XMM-Newton during the respective time period. 

To assess the significance of these detections, we proceed with the same $\Delta$AIC methodology as outlined above. However, stacking the many individual PSDs for NuSTAR allows us to assume Gaussian error bars. Thus, we adopt the standard Gaussian likelihood given by
\begin{equation} \label{eqn:gauss_like}
    \ln \mathcal{L} = -\frac{1}{2} \sum_i \left[\left(\frac{x_i - \mu_i}{\sigma_i}\right)^2 + \ln(2\pi\sigma_i^2)\right],
\end{equation}
where $i$ denotes a sum over all frequencies, $x_i$ is the power at a frequency $f_i$, $\mu_i$ is the model power at a frequency $f_i$, and $\sigma_i$ is the Gaussian uncertainty at a frequency $f_i$. We test the presence of a QPO in the NuSTAR data by fitting both a Power-Law $+$ Constant model and a Power-Law $+$ Constant $+$ Lorentzian model to the data. In all four stacked PSDs, the additional Lorentzian provides a statistically significant improvement over the Power-Law $+$ Constant model. In all detections, the QPO frequency is formally consistent with the XMM-Newton results (as seen in Figure \ref{fig:qpo_freq_evol}), but with significantly more uncertainty on the width and RMS, due to both averaging over significant frequency evolution (in the 2023 data) and the worse frequency resolution of these short NuSTAR stacked PSDs. By stacking all data from 2024-2026, when the QPO frequency was roughly constant, we can achieve a 6$\sigma$ overall detection in the 3-10 keV NuSTAR data, making us highly confident in this QPO detection. The count rates are extraordinarily low above 10 keV, making a firm detection of the QPO outside of the XMM-Newton bandpass challenging. We are able to detect the QPO in 5-30 keV NuSTAR data at roughly 4$\sigma$ significance, but the RMS of this feature is consistent with the measurement in the 3-10 keV band with NuSTAR. Hence, we are unable to confidently say whether the QPO continues to strengthen with increasing energy above 10 keV. 

\subsection{Evolution of QPO and Broadband Noise Properties}

\begin{figure}
    \centering
    \includegraphics[width=\linewidth]{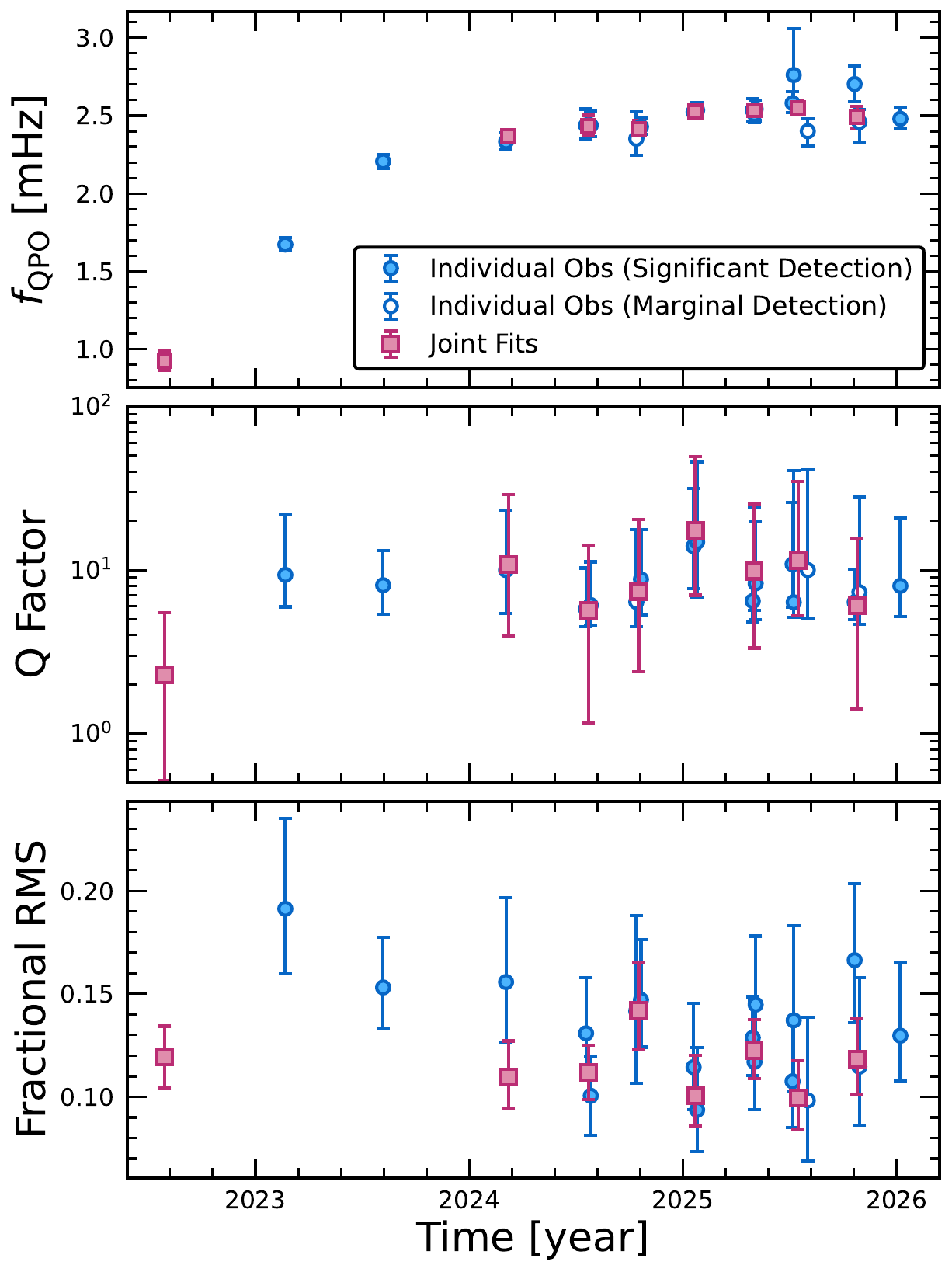}
    \caption{Evolution of the QPO properties with time, where the color coding and shapes match Figure \ref{fig:qpo_freq_evol}. \textit{Top:} Evolution of the QPO frequency with time, as plotted in Figure \ref{fig:qpo_freq_evol}. \textit{Middle:} Evolution of $Q = f_\mathrm{QPO}/\mathrm{FWHM}_\mathrm{QPO}$. Besides the observations in 2022, which showed a broad QPO, there is no significant evolution in the width of the QPO. Performing joint fits does not significantly improve the measurement of $Q$ as we are more so limited by the number of frequency bins over which we see the signal. \textit{Bottom:} Evolution of the fractional RMS of the QPO, as measured from the normalization of the fitted Lorentzian component. The RMS remains stable in the recent observations, after a potential increase in 2023. We can significantly improve the uncertainty on the RMS by jointly fitting multiple observations in a given epoch. }
    \label{fig:QPO-props}
\end{figure}

With our multi-year XMM-Newton campaign, we also searched for significant evolution in the other QPO properties, detecting no significant evolution in the 2024-2026 observations. Figure \ref{fig:QPO-props} shows the evolution of the $Q$ factor and fractional RMS (as measured from the normalization of the Lorentzian fit to the PSD) over the course of this monitoring campaign. No significant evolution is seen in either parameter in the 2024-2026 data; the QPO is remaining remarkably stable, despite significant evolution in the QPO frequency at early times. 

\subsection{A Search for Harmonics and Breaks in Stacked PSDs}

\begin{figure}
    \centering
    \includegraphics[width=\linewidth]{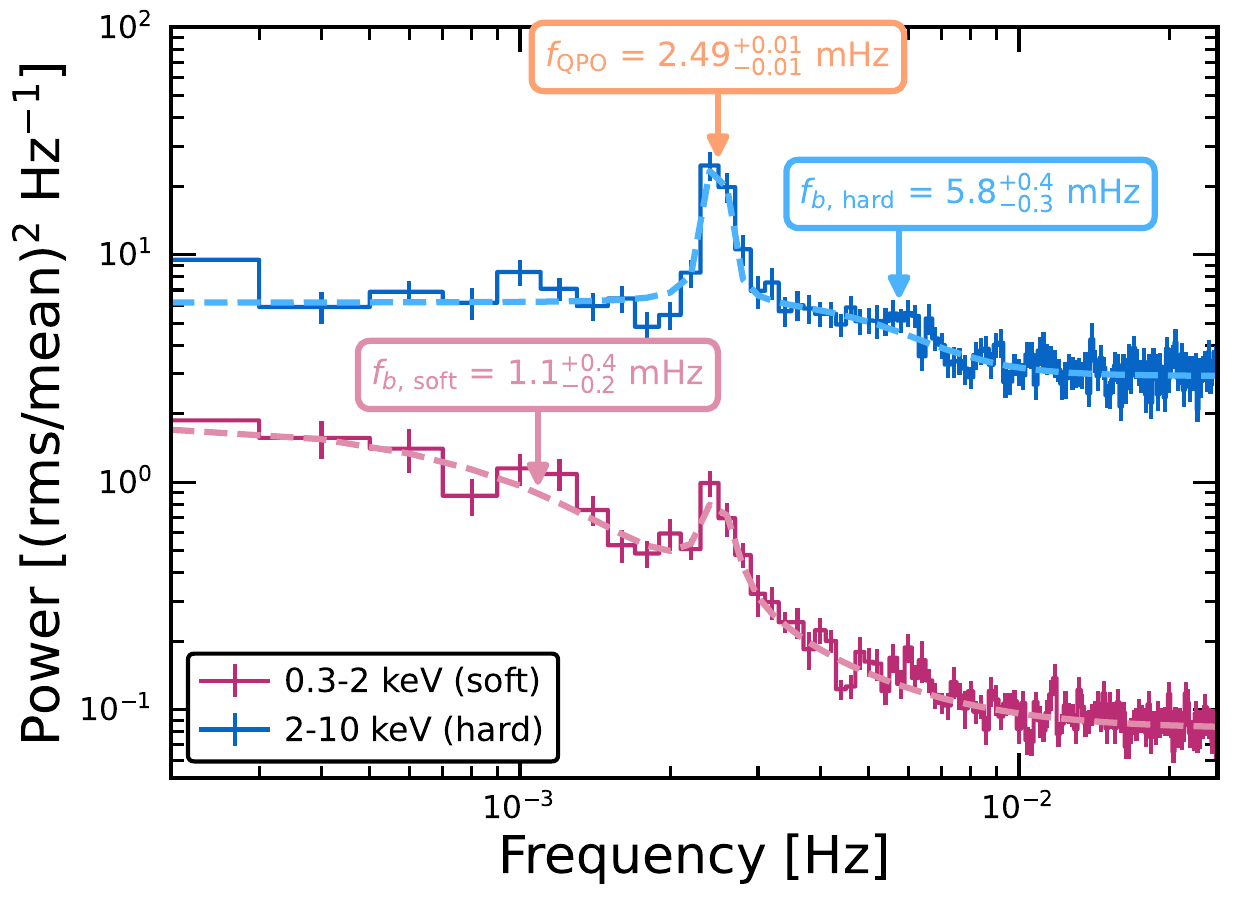}
    \caption{Stacked PSD of all XMM-Newton observations from July 2024-January 2026 in the 2-10 keV band (blue) and 0.3-2 keV band (pink). Each of the stacked PSDs are fit with a broken power-law broadband noise model and a Lorentzian for the QPO, the best fit of which are shown as dashed lines. The orange arrow shows the the fundamental QPO frequency as measured from the 2-10 keV PSD, and the blue and pink arrows show the break frequencies of their respective bands. No significant second harmonic is detected when we consider a broken power-law broadband noise model.}
    \label{fig:stackedPSD_XMM}
\end{figure}

With the 16 XMM-Newton observations from July 2024-January 2026 with any level of QPO detection, we see negligible evolution in the QPO frequency and can therefore combine the PSDs to search for the existence of additional harmonic features and breaks in the broadband PSD slope that may be too weak to detect in a single observation alone. To do this, we proceed in a similar manner to the NuSTAR analysis; we break each XMM-Newton observation into as many continuous 5~ks segments as possible, compute one PSD per segment, and stack the resulting PSDs together. In these 16 observations, we have a total of 51 segments, putting us again firmly in the Gaussian regime (i.e., to fit the stacked PSD, we use the likelihood given by Equation \ref{eqn:gauss_like}). Figure \ref{fig:stackedPSD_XMM} shows the resulting PSD in both the 2-10 keV (blue) and 0.3-2 keV (pink) bands. The location of the QPO is highlighted in orange ($f_\mathrm{QPO} = 2.49^{+0.01}_{-0.01}$ mHz). With 51 $\times$ 5~ks and the measured QPO frequency, we detect roughly 630 QPO cycles in these 16 observations alone. 

With this exquisite data set and large number of QPO cycles, we can begin to probe the PSD for weaker features, like a second harmonic and breaks in the broadband noise. There is some structure in the stacked PSD above the QPO frequency, particularly in the 2-10 keV band. To test the statistical significance of this structure, we fit the data with a variety of models, including: 
\begin{enumerate}[(1)]
    \item Power-law + 1 Lorentzian
    \item Power-law + 2 Lorentzians
    \item Broken power-law + 1 Lorentzian
    \item Broken power-law + 2 Lorentzians
\end{enumerate}
In Models (3) and (4), we fix the low-frequency slope of the broken power-law to $\alpha = 0$, based on the shape of the PSD and the relatively low number of points at low frequency. The second Lorentzians in Models (2) and (4) are intended to fit a potential second harmonic in the PSD. Model (2) provides a significant improvement over Model (1) with $\Delta$AIC $\approx$ 27 ($4.8\sigma$), but the additional Lorentzian is quite broad ($Q \approx 1.5$) and is at a non-integer frequency ratio with the fundamental ($f_\mathrm{Lor,\,1}/f_\mathrm{Lor,\,2} \approx 1.75$). Model (3) provides a similar improvement over Model (1) with $\Delta$AIC $\approx$ 24 ($4.5\sigma$), suggesting that a break in the PSD and a broad harmonic provide a statistically equivalent fit. It is worth noting that BHXRB PSD modeling has also found that multiple Lorentzians provide a better fit to the broadband noise than broken power-law models \citep[e.g.,][]{Nowak2000,Belloni2002}, suggesting that such a degeneracy between a broad Lorentzian and a break in the power-law model is unsurprising. Finally, Model (4) provides no improvement to Model (3), and thus there is no clear evidence for a second harmonic feature in the stacked PSD.

The need for a break in the broadband noise is far more significant in the soft band, for which Model (3) provides a significant improvement over Model (1), with $\Delta$AIC $\approx$ 81 ($8.7\sigma$). For both bands, the resulting best fit for Model (3) is shown for each model as a dashed line in Figure \ref{fig:stackedPSD_XMM}. The break frequency differs significantly between the two bands with $f_{b,\, \mathrm{hard}} = 5.8_{-0.3}^{+0.4}$~mHz and $f_{b,\,\mathrm{soft}} = 1.1_{-0.2}^{+0.4}$~mHz. Similar behavior has been suggested in a handful of individual AGN without QPOs \citep[e.g.,][]{McHardy2004,Alston2019,Ashton2021}, but there are significant degeneracies between the break frequency and high-frequency slopes, especially in single observations. In the propagating fluctuations model, an increasing break frequency (or high energy slope) with increasing energy can be explained by different emissivity indices in different energy bands \citep[i.e., a radially stratified corona;][]{Arevalo2006,Ashton2022}. Using previously reported relationships between the PSD break frequency and black hole mass \citep[e.g.,][]{McHardy2006,Gonzalez-Martin2012}, we find that the soft-band break frequency is consistent with the $M \approx 10^6\, M_\odot$ measurement for the black hole in \src\, \citep{Li2022}.  

\subsection{Lag Spectra}

Fourier-resolved time lags between different energy bands can provide unique insights into the driving mechanism of the variability in accreting black holes \citep[see, e.g.,][]{Uttley2014,Kara2016,Cackett2021,Wang2022}. We compute lags in \src\, using the STELA Toolkit\footnote{\url{https://collinlewin.github.io/STELA-Toolkit/}}, which follows the standard Fourier timing techniques outlined in \cite{Uttley2014}. In brief, we compute the Fourier transform of light curves ($\Delta t = 10$s) in two different energy bands and multiply one by the complex conjugate of the other to produce the cross-spectrum. The time lag between these two bands is then computed via $\tau(f) = \phi(f)/2\pi f$, where $\phi(f)$ is the frequency-dependent phase of the cross-spectrum. We adopt the standard convention that a positive lag corresponds to a hard lag (i.e., the harder energy band lags behind the softer band). 

\begin{figure*}
    \centering
    \includegraphics[width=\linewidth]{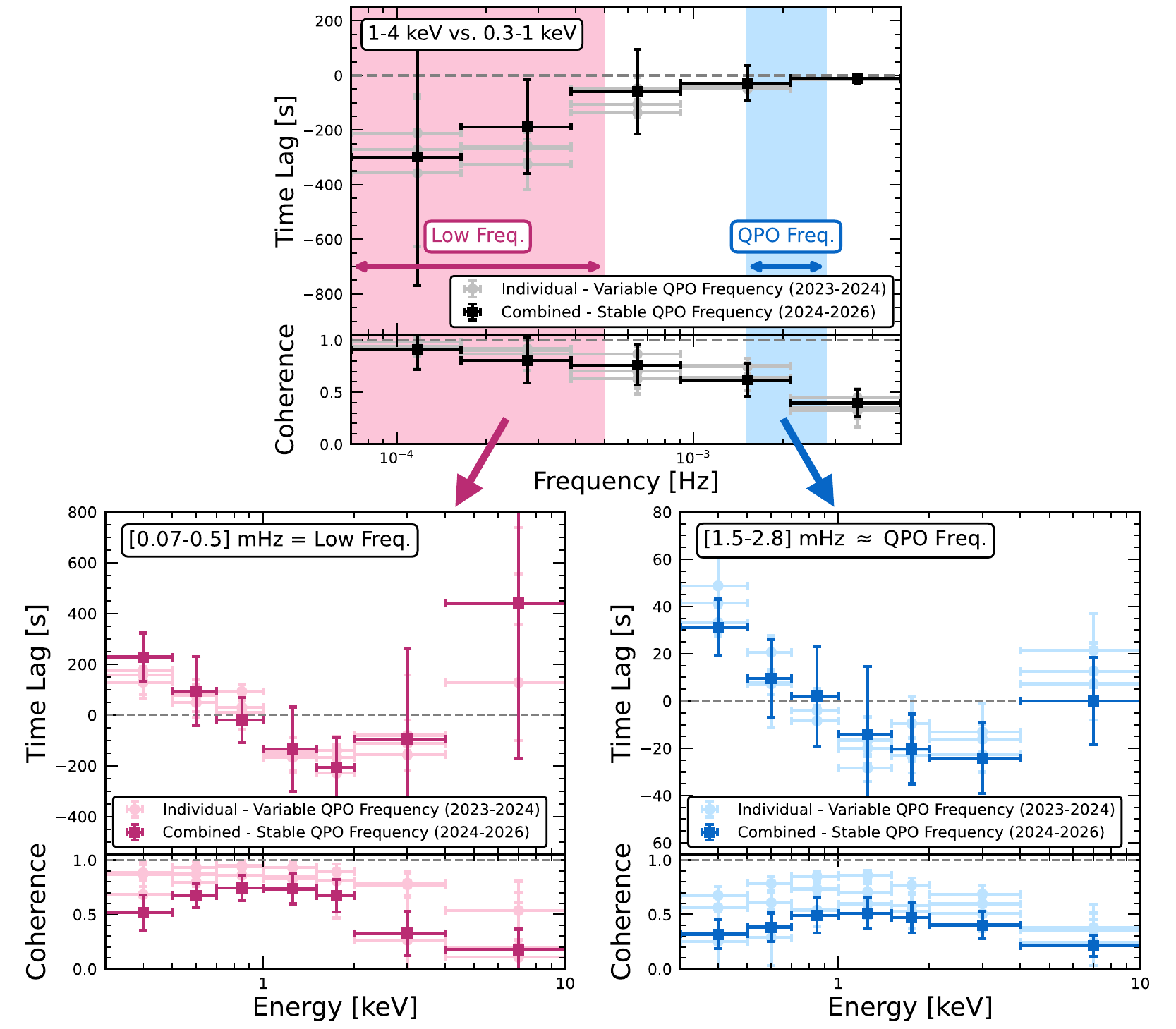}
    \caption{Lag spectra during the QPO phase of \src, constructed with 10s binned light curves and adopting the convention that a positive lag indicates that the hard band lags behind the soft band. {\it Top panel:} LFS between the 0.3-1 keV and 1-4 keV bands. The gray points show the individual LFS measurements for the observations in which the QPO frequency significantly changed (i.e., those observations presented in \citetalias{Masterson2025}), while the black points show the mean of the measurements for which the QPO frequency was stable (i.e., all new observations presented in this work). For the averaged LFS and coherence, the error bars represent the standard deviation to highlight the spread of values. The pink and blue regions denote the frequency regimes in which were compute the LES, with pink corresponding to the low frequency regime (0.07-0.5 mHz) and the blue corresponding to the QPO frequency regime (1.5-2.8 mHz). {\it Bottom panels:} LES in two different frequency regimes (left/pink = low frequency, right/blue = QPO frequency). As in the top panel, the lighter shades show the individual measurements from when the QPO frequency evolved, whereas the darker shades show the averaged measurements from when the QPO frequency was stable. We use the 0.3-10 keV band as the reference band, subtracting off the band of interest from each lag measurement to avoid correlated noise. The two LES show similar shapes --- rising towards both low and high energies --- but different amplitudes. The shape of each LES is consistent among all observations, suggesting a common driver of the variability, despite the significant change to the QPO frequency.}
    \label{fig:LFS-LES}
\end{figure*}

We first compute a lag-frequency spectrum (LFS) for each observation between the 0.3-1 keV band and 1-4 keV band, using a consistent binning scheme across all observations of five logarithmically-spaced bins from $7 \times 10^{-5}$ Hz to $5 \times 10^{-3}$ Hz (above which the coherence drops significantly). These energy bands were chosen to allow for a direct comparison to previous AGN lag analyses \citep[e.g.,][]{Fabian2009,DeMarco2013,Kara2016}; the 1-4 keV band is dominated by continuum power-law emission from the corona, while the 0.3-1 keV band is dominated by the soft excess. Hence, soft (negative) lags are often indicative of reflection, while hard (positive) lags have been suggested to be related to inward-propagating fluctuations. In addition to the observations presented in Table \ref{tab:obs}, we also include ObsIDs 0915390701, 0931791401, and 0932392001 from \citetalias{Masterson2025} in this analysis. We exclude the data from ObsID 0953010901 from this analysis, as the cleaned data does not contain a sufficient baseline to compute lower frequency lags. Most observations show a consistent LFS shape within uncertainties. To highlight the lack of evolution despite the significant change to the QPO frequency, we show the individual measurements of the LFS from the three ObsIDs from \citetalias{Masterson2025} compared to an averaged LFS using all new observations at a roughly constant QPO frequency in Figure \ref{fig:LFS-LES}. 

\begin{figure}
    \centering
    \includegraphics[width=\linewidth]{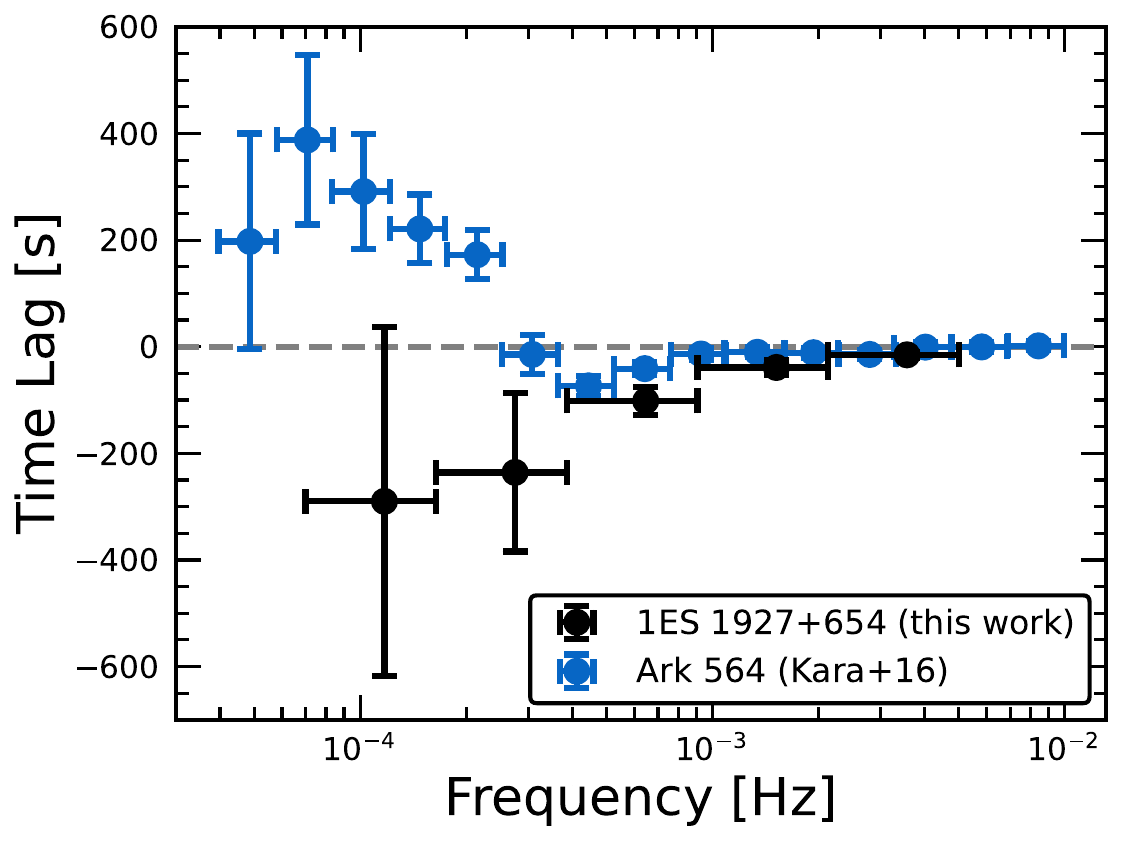}
    \caption{Comparison of the LFS in \src\, (black) with all observations from 2023-2026 to that of Ark 564 (blue) from Figure 3 of \cite{Kara2016}. Ark 564 has both a similar mass ($\log (M_\mathrm{BH}/M_\odot) \approx 6$) and mass accretion rate ($L/L_\mathrm{Edd} \approx 1$) as \src. \src\, does not show the low frequency hard (positive) lags as in Ark 564, although we note that the short observations of \src\, limit our ability to constrain the lowest frequencies to high precision.}
    \label{fig:ark564_comp}
\end{figure}

The LFS shows a soft (negative) lag across all measurable frequencies. Many AGN show a soft lag commonly associated with reverberation at high frequencies, with a reversal to a hard lag commonly associated with mass accretion rate fluctuations at low frequencies \citep[e.g.,][]{DeMarco2013,Kara2016}. One possibility to explain the persistent soft lag in \src\, is that we have not probed low enough frequencies to see the turn over in the LFS. The frequency of the soft lag in AGN correlates inversely with the black hole mass and correlates positively with mass accretion rate \citep{Kara2016}. In Figure \ref{fig:ark564_comp}, we show a comparison between the LFS in \src\, and that of Ark 564 which has a similarly low mass ($M_\mathrm{BH} \approx 10^6 \, M_\odot$) and high accretion rate ($L/L_\mathrm{Edd} \approx 1$) to \src. We note that the current X-ray luminosity of \src\, is $L_X / L_\mathrm{Edd} \approx 0.5$, justifying this high accretion rate comparison. While Ark 564 shows clear evidence for a low-frequency hard (positive) lag, \src\, does not (albeit with limited precision of the lowest frequency lags due to short observation durations). The lack of this low-frequency hard (positive) lag could suggest that the canonical accretion rate fluctuations are intrinsically not present in \src. It is interesting to note that beginning in 2022, the X-ray and UV variability began to decouple in this source \citep{Ghosh2023,Laha2025}. Although this behavior is on timescales longer than what we probe here, this lends credence to the lack of propagating fluctuations in \src.

To investigate the variability further, we construct lag-energy spectra (LES) in different frequency ranges. The bottom two panels of Figure \ref{fig:LFS-LES} show the resulting LES in the low-frequency regime (0.07-0.5 mHz, left/pink) and near the QPO frequency (1.5-2.8 mHz, right/blue). The lighter points show the individual LES from the three observations with clear QPO frequency evolution. As with the LFS, the darker points show the stacked LES for the 2024-2026 data with constant QPO frequency. The lag amplitude and shape of the LES do not evolve throughout the observations analyzed in this work, despite the significant change to the QPO frequency in the early data. The LES is similar in both the low frequency and the QPO frequency range, with only the amplitude of the lag changing significantly between these two frequency ranges. The shape of the LES is akin to what is seen in the high-frequency reverberation lags in AGN, although the coherence is quite low in the highest energy band (4-10 keV) leading to large uncertainties on that lag measurement. Recent NuSTAR observations have revealed a possible broad iron line between 6-7 keV appearing from 2022-2025 (D. Sadaula et al. submitted), potentially allowing us to connect the LES to canonical reverberation in AGN, although with caveats that are explored in Section \ref{sec:discussion}. 

\subsection{Extreme X-ray Jumps on the QPO Period} \label{subsec:jumps}

\begin{figure*}
    \centering
    \includegraphics[width=\textwidth]{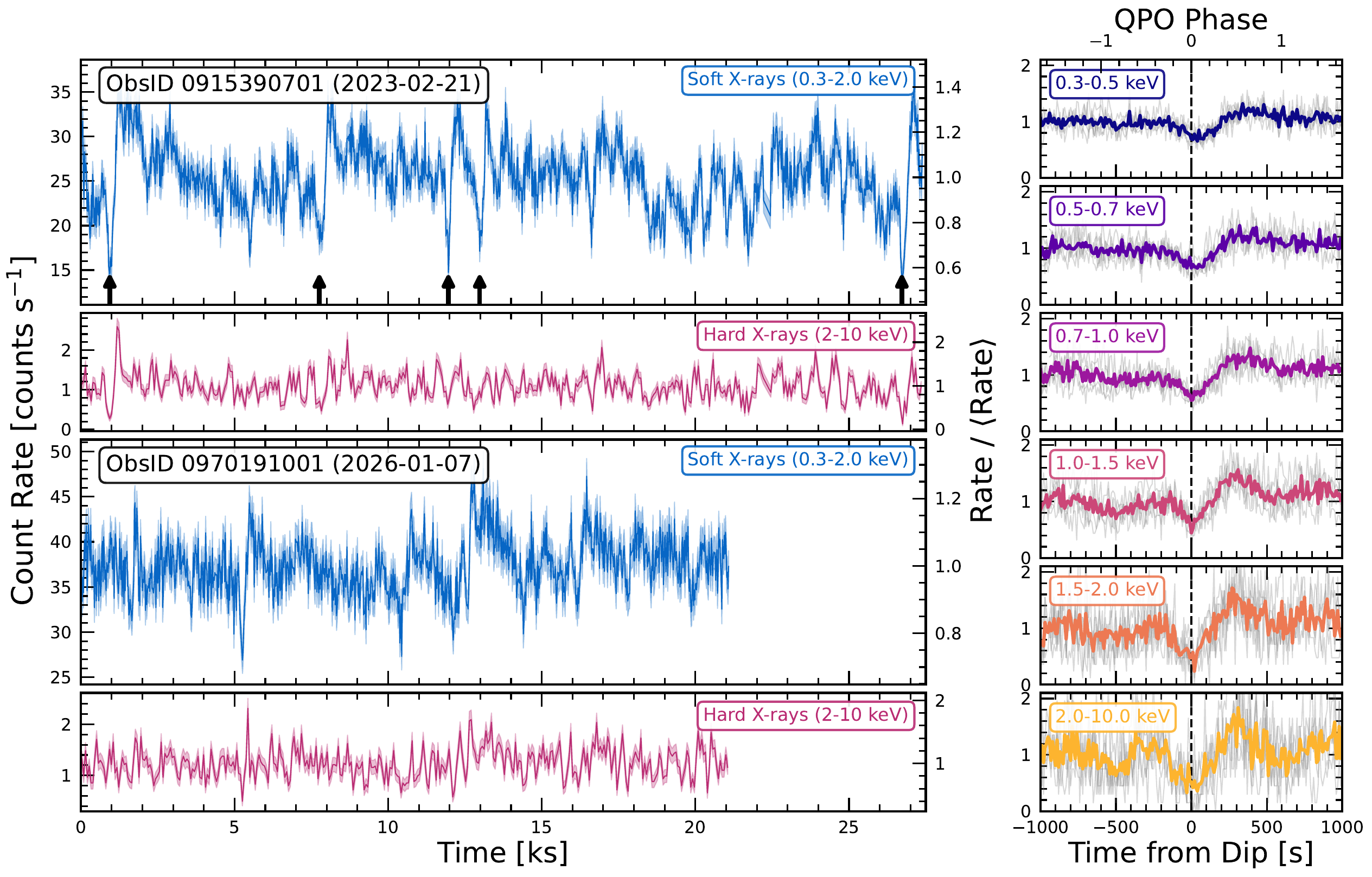}
    \caption{Strong X-ray jumps on the QPO period in \src. \textit{Left:} 0.3-2 keV (blue) and 2-10 keV (pink) light curves for the observation from February 2023 with a 1.67 mHz QPO (top) and from January 2026 with a 2.48 mHz QPO (bottom). Both observations show repetitive sudden drops in the X-ray flux by $>20\%$, followed by a rapid increase of the X-ray flux. These occur on the QPO timescale, but do not repeat as regularly as the oscillations seen in the hard X-ray band. \textit{Right:} Jump profiles in the February 2023 observation as a function of energy for the five most extreme jumps highlighted with pink arrows in the top left panel. The zero time was chosen to be the minimum time in the 0.3-2 keV band shown to the left. The gray lines show the individual jumps, while the color lines show the average over the five jumps. The y-axis limits are the same for all energy ranges to highlight that the jumps are strongest (in regards to fractional variability) in the 2-10 keV band. From these panels, we can also see that the lower energy bands slightly lead the medium energy bands, consistent with the LES shown in Figure \ref{fig:LFS-LES}.}
    \label{fig:jumps}
\end{figure*}

One of the most striking features in the XMM-Newton observations of \src\, are repetitive, major ``jumps" in the total X-ray flux. The left panels of Figure \ref{fig:jumps} show light curves from February 2023, when the jumping was first apparent, and January 2026, the latest epoch analyzed in this work. The right panels of Figure \ref{fig:jumps} show the energy dependence for these jumps in the February 2023 observation. For this panel, the gray lines show the individual jumps, while the colored lines mark the average; for this panel, we use the five strongest jumps that are marked with black arrows in the left panel which were selected based on extreme peaks in the derivative of the light curve (smoothed over 60s intervals to mitigate Poisson fluctuations). The top axis shows the QPO phase, highlighting that this behavior is inherently related to the QPO, as it occurs on the same timescale and at a fixed phase. The same holds for the more recent observations, which, together with the fact that we see these jumps in nearly all observations that we detect a hard X-ray QPO in, strengthens the connection between these jumps and the hard X-ray QPO. 

\begin{figure}
    \centering
    \includegraphics[width=\linewidth]{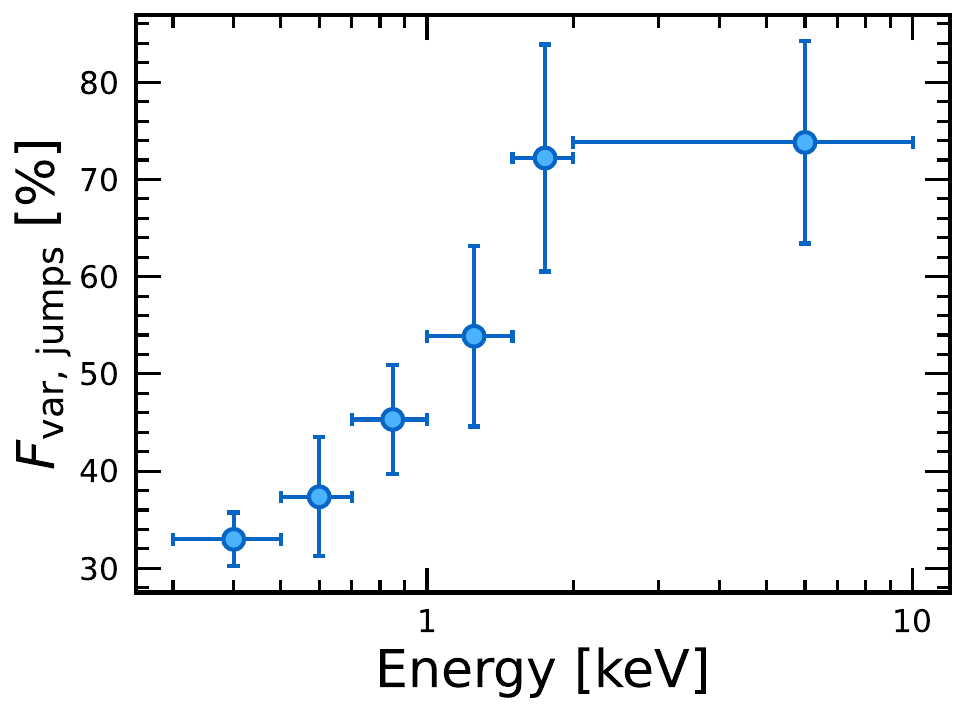}
    \caption{Fractional variability of the 5 major jumps shown in the right panel of Figure \ref{fig:jumps} (from February 2023), where error bars denote the standard deviation across the 5 different jumps. These values were computed in the time-domain by measuring half the difference from the maximum to minimum normalized rates. The energy-dependent shape of the jump amplitude is comparable to that of the QPO (see \citetalias{Masterson2025} and \citealt{Shui2025}), but with significantly higher fractional variability. }
    \label{fig:jumps-fvar}
\end{figure}

While we use the term ``jump" to describe these features, we stress that the phenomenology is more rich than a single jump. Namely, the jumps always first show a dip in the X-ray flux, then a rapid rise to higher than the pre-dip flux, and finally a decline back to its pre-dip flux. The shape and phase of the jumps is nearly identical among all events and all observations. The same shape is broadly seen across all X-ray energies, but Figure \ref{fig:jumps-fvar} shows that the fractional variability rises with increasing energy, peaking in the 2-10 keV band. We note that this behavior is similar in shape to what is seen in the QPO energy dependence (\citetalias{Masterson2025}, \citealt{Shui2025}), but with a much larger amplitude. The steep energy dependence effectively rules out transient obscuration effects that block the entire X-ray source and driving dips, as is commonly invoked in BH and neutron star XRBs where the X-ray spectrum hardens during the dips \citep[e.g.,][]{Kuulkers1998,DiazTrigo2006}. Finally, each individual jump is highly coherent, but there appears to be no apparent periodicity between the jumps. Together, these observables can be used to constrain models for the jumps, which we explore and attempt to connect to the origin of the QPO in Section \ref{sec:discussion}. 

\section{Discussion} \label{sec:discussion}

With the new XMM-Newton observations presented in this work, we have reached over 900 observed QPO cycles in \src, making this one of the most exquisite data sets to test the underlying physics driving this QPO and connect it to both BHXRB QPOs and other quasi-periodicity in SMBHs like quasi-periodic eruptions \citep[QPEs;][]{Miniutti2019,Giustini2020,Arcodia2021,Chakraborty2021}. In this section, we discuss the implications for various timing analyses reported in Section \ref{sec:results} on the QPO mechanism and the underlying accretion flow in this enigmatic AGN. 

\subsection{QPO Frequency Evolution}

An important result of this work is that the QPO frequency has stalled at a frequency of around 2.5 mHz, showing no strong evidence for evolution in the last 1.5 years. This is in stark contrast from the significant increase in QPO frequency seen between 2022-2024, which made this QPO particularly unique compared to other AGN and HFQPOs in BHXRBs. \citetalias{Masterson2025} suggested that the evolution in the QPO frequency could be from an oscillating, contracting corona or an orbiting, mass-transferring companion. The stabilization of the QPO frequency does not alone break degeneracies between these two models, but it suggests that either the corona size and temperature have stabilized or the companion has reached a point where the angular momentum losses from GR balance the angular momentum gains from mass transfer (i.e., -$\dot{J}_\mathrm{GR} = \dot{J}_\mathrm{MT}$). Another alternative to these models relates to the inner edge of the accretion disk; if the disk is truncated, it can evolve inward roughly on the viscous timescale, stabilizing when it reaches the ISCO. At $10\, R_g$, the viscous timescale can match the $\sim 2$ year evolution with reasonable choices of $\alpha$ and disk scale height (e.g., $\alpha \sim 0.01$, $h/r \sim 0.01$). However, we note that the viscous evolution of the inner edge of the disk should produce a runaway effect, getting faster as the disk propagates inward. Additional effects (e.g., from magnetic fields) would be needed to stall such propagation to match the observed evolution.

The other AGN with a similarly strong and persistent QPO is RE J1034+396 \citep{Gierlinski2008,Middleton2009,Alston2014,Taylor2025}, which, unlike \src, shows no secular evolution in its QPO frequency. However, it does does show $\sim10\%$ stochastic variability around the central QPO frequency \citep{Xia2024}, as has been seen in HFQPOs in BHXRBs \citep{Belloni2012}. One speculative question that is worth asking is: is \src\, is transitioning into a state more akin to what is seen in RE J1034+396 and BHXRB HFQPOs as it has plateaued? In the 1.5 years of plateau presented in this work, we see very mild frequency jitter in the QPO in \src\, during the plateau phase of $\sim 5\%$, albeit with comparable uncertainty levels in individual measurements. Continued monitoring of the QPO in \src\, will help elucidate this question, as persistent constant frequency would lend credence to this comparison, while revamped secular evolution would support a unique origin for the QPO in \src.

\subsection{Models for Extreme X-ray Jumps}

The dips seen at the start of the extreme X-ray jumps in \src\, are reminiscent of eclipses, which are commonly seen in compact and stellar binaries. For example, $\eta$ Carinae, a massive star with a companion on an eccentric orbit that produces wind-wind collision shocks at pericenter \citep[e.g.,][]{Pittard2002,Corcoran2005}, shows similar X-ray dips and flares to what we see in \src. The X-ray light curve of $\eta$ Carinae peaks first, followed by a rapid dip likely caused by the shocked emission being blocked by the dense stellar wind of the primary star \citep{Hamaguchi2014,Panagiotou2018}. Accreting compact binaries have shown similar behavior; for example, in double white dwarf systems undergoing direct impact accretion, the accretion stream can be deflected by the Coriolis force, leading to the rise from hot spot emission produced by the impact preceding the dip from the eclipse \citep[e.g.,][]{Marsh2002,Barros2007}. To reproduce the dip seen \textit{before} the rapid rise in \src, we would need to see an emission mechanism that trails rather leads the eclipse. The emission also needs to be set by the same orbital period as the eclipses to produce such linked behavior. Self-lensing by a companion \citep[e.g.,][]{Ingram2021} naturally gives rise to a linked period, and recent work has shown that the flares can be more asymmetric than initially thought \citep{Davelaar2022,Krauth2024}. Likewise, shocks, akin to some models for QPEs \citep{Linial2023,Franchini2023}, could be delayed due to photon diffusion and ejecta expansion timescales \citep[see e.g.,][]{Huang2025}. 

In this eclipsing model, and in light of the suggestion of a white dwarf companion \citepalias{Masterson2025}, it is prudent to ask whether a white dwarf is capable of eclipsing the X-ray emitting region near the SMBH (from which most of the luminosity we see is expected). An eclipse can occur if the emitting region is significantly smaller than the eclipsing body, and the dip depth should scale as the area of the companion over the emitting region area (i.e., $\delta \propto (R_\mathrm{WD}/R_\mathrm{corona})^2$). For a 0.1 $M_\odot$ white dwarf, the approximate donor mass at which Roche-lobe overflow would occur at the QPO period as proposed in \citetalias{Masterson2025}, the radius is $R_\mathrm{WD} \approx 10^9$~cm $\approx 0.01 \, R_g$ for a $10^6 \, M_\odot$ SMBH companion (assuming a zero-temperature white dwarf mass-radius relation). Thus, it is nearly impossible to produce such strong X-ray dips ($\gtrsim 20\%$), as decreasing the corona size beyond $\sim$ few $R_g$ will lead to runaway pair-production \citep{Fabian2015}. However, the white dwarf itself need not be the eclipsing body; in QPEs, the expanding shocked ejecta is often claimed to be optically thick at early times with a radius on the order of $10^{11}$~cm \citep[e.g.,][]{Chakraborty2024}. This corresponds to roughly the gravitational radius for a $10^6 \, M_\odot$ SMBH and thus could sufficiently block the hard X-ray emitting region, although it remains unclear whether the same emission mechanisms can hold for a white dwarf companion. In this picture, only the near-side ejecta would be able to block our line-of-sight, potentially explaining why the jumps are somewhat sporadic, while the overall QPO appears stable. 

It is also worth exploring models that invoke mass accretion rate changes to explain the jumping behavior. Extreme changes to the mass accretion rate can be driven by a variety of factors, including disk tearing instabilities, which lead to the rapid depletion of the innermost regions of the accretion flow as the innermost ring rapidly accretes \citep{Nixon2012,Raj2021}. Such extreme changes to the disk flux should in turn affect the corona, as the coronal properties depend sensitively on the cooling rate set by properties of the seed photons. Existing theoretical work has shown that limit cycles between the disk and corona properties can qualitatively explain low-frequency QPOs in BHXRBs \citep[e.g.,][]{Lopez-Barquero2025}. However, these models are at odds with the observed jumps in \src\, in two key ways. First, such coupling would predict that the X-ray spectrum will be softer (i.e., larger $\Gamma$, lower $kT_e$) when the flux is higher, akin to what is observed in most AGN \citep{Markowitz2003a,Sobolewska2009,Connolly2016} but opposite what is seen in \src. Second, it is difficult to reconcile these ideas with the dips preceding the flares, as the X-ray flux should roughly track the accretion rate and input seed photon flux. Thus, while it is appealing to connect major changes in the disk to both the jumps and oscillations in the corona, existing models cannot properly reproduce the observed energy dependencies. 

Finally, it is worth comparing to ultraluminous X-ray sources (ULXs), which are also thought to be rapidly accreting above the Eddington limit \citep{Kaaret2017}. In particular, the ULX NGC 247 X-1 has shown repetitive quasi-periodic X-ray dips that also increase in strength as a function of energy \citep{Alston2021}. The presence of strong absorption features in the spectrum strengthen the interpretation of these dips as an optically thick structure occulting the small X-ray emitting region through either a warped disk, wind, or combination of the two \citep{Pinto2021}. However, these dips occur on a much longer timescale when accounting for the mass difference and no obvious absorption features are seen in the X-ray spectrum of \src, although future work will investigate this in more detail through phase-resolved spectroscopy before, during, and after the jumps.

\subsection{Origin of the Soft X-ray Lag}

The lags in \src\, can provide valuable insights into the geometry and emission mechanisms in the inner accretion flow. Broadly speaking, the shape of the LES in \src\, is consistent with reverberation in AGN with a rise of the lag in the Fe K band. We do not have the spectral-timing resolution to say definitively that this is reverberation, but the recent detection of an Fe K line (D. Sadaula et al. submitted) and prior reflection model for the transient 1 keV line \citep{Masterson2022} strengthen this potential model. The lag at soft energies could also be related to reverberation, if the soft excess is produced by relativistically blurred reflection \citep[e.g.,][]{Crummy2006}. Similarly, \cite{Shui2025} recently presented a similar LES of \src\, at the QPO frequency with different phase-resolved methods. Broadly their LES matches the one that we see using Fourier techniques at the QPO frequency and can be modeled using the time-dependent Comptonization model, \texttt{vKompth} \citep{Bellavita2022}, that includes feedback between the disk and corona akin to reflection.

If we are seeing a reverberation lag, then it is puzzling for two key reasons: (1) the feature persists to lower frequencies than expected for AGN of similar mass (see the comparison to Ark 564 in Figure \ref{fig:ark564_comp}), and (2) the lag amplitude at low frequencies is large for a typical $10^6~M_\odot$ black hole according to scaling relations presented in \cite{Kara2016}, albeit with large uncertainties due to the short observations of \src. To the first point, at low frequencies ($f \lesssim 10^{-4}$ Hz), nearly all AGN show hard lags and a log-linear increase in the LES to high energies \citep{Kara2016}. These low-frequency hard lags were first discovered in BHXRBs and have long been associated with propagating fluctuations in the accretion flow \citep[e.g.,][]{Lyubarskii1997,Kotov2001,Arevalo2006}. While we defer a detailed analysis of the RMS-flux relationship and the lack of propagating fluctuations to subsequent work, it is interesting to note that \src\, also clearly breaks the canonical RMS-flux relation of accreting sources that also is thought to arise from propagating fluctuations \citep{Uttley2001,Uttley2005}, showing very little variability at its peak X-ray flux (see e.g., the peak of the NICER light curve from mid-2019 to mid-2020 in \citetalias{Masterson2025}). 

The other puzzling aspect of the lags is that they do not evolve significantly, despite significant evolution in the QPO frequency. In BHXRBs, the reverberation lag has been shown to evolve significantly over the course of an outburst, namely showing a shift towards higher frequencies and lower amplitudes during the rise to the luminous hard state \citep{Kara2019}, followed by a shift towards lower frequencies and higher amplitudes during the state transition \citep{Wang2021,Wang2022}. Both of these behaviors have been interpreted as correlating with the extent of the corona, the former through contraction of the corona, and the latter through expansion, potentially related to transient jet ejections \citep[although recent X-ray polarization measurements are challenging this picture; e.g.,][]{Ingram2024}. In \src, if the change to the QPO frequency is primarily driven by changes to the size of the corona \citepalias[one option in the coronal oscillations model;][]{Masterson2025}, then we would expect to see a change either to the lag amplitude or frequency of the soft lag, neither of which are seen. Thus, we suggest that the QPO frequency evolution is driven either by changes to the corona temperature or another physical mechanism altogether.

Low-frequency soft lags have also been observed in ULXs \citep[e.g.,][]{Heil2010,Pinto2017,Kara2020} and inferred to exist in RE J1034+396 \citep{Taylor2025}, the other AGN with a similarly strong and persistent X-ray QPO \citep{Gierlinski2008}. In RE J1034+396, the lag was shown to undergo reversals correlated with the QPO frequency \citep{Xia2024}, which has recently been interpreted as a persistent soft lag of $\sim2000$s that changes sign due to phase wrapping at roughly the QPO period. These intrinsic soft lags in both ULXs and RE J1034+396 have been interpreted as Compton scattering of the coronal photons either in a thick outflow or the accretion disk \citep{Kara2020,Taylor2025}. In ULXs, this picture is complicated by a large lag amplitude implying extremely high outflow densities \citep{Kara2020}, which does not exist in \src\, given the the larger physical scales and comparable lag amplitude. Hence, Comptonization is a viable mechanism for producing the soft lag at low energies. 

\section{Conclusions} \label{sec:conclusion}

In this work, we report the continued detection of the mHz X-ray QPO in \src\, and subsequent X-ray timing analyses, including X-ray lag spectra, deep stacked PSDs, and dipping behavior. A full theoretical picture for this source remains elusive, and this work has opened up new questions relating to the origin of the strong, phase-locked dips at the QPO timescale. Our main findings are: 

\begin{itemize}
    \item The QPO in \src\, persists to present day and has plateaued around a frequency of $f_\mathrm{qpo} \approx 2.5$ mHz, more than double the initial detection in \citetalias{Masterson2025} at $f_\mathrm{qpo} \approx 0.9$ mHz (left panel of Figure \ref{fig:qpo_freq_evol}). The QPO frequency continues to roughly track the X-ray flux, although we see some stochasticity in this relationship (right panel of Figure \ref{fig:qpo_freq_evol}).
    \item We detected the QPO in a stacked PSD from four epochs of NuSTAR observations from 2023 to 2026 (Figure \ref{fig:nustar_psd_all}). The frequencies of the QPO in NuSTAR matched that of the XMM-Newton data at the same time. To our knowledge, this is the first detection of an SMBH QPO with NuSTAR, thereby opening up a new discovery window. 
    \item By stacking 16 XMM-Newton observations with minimal QPO frequency evolution, we were able to exclude a second harmonic feature at standard harmonic ratios (e.g., 3:2, 2:1, 3:1). This stacked PSD also revealed strong evidence for a break in both the soft and hard X-ray bands, but with significantly different frequencies in the two bands (Figure \ref{fig:stackedPSD_XMM}). 
    \item The lag-frequency spectra calculated with 1-4 keV versus 0.3-1 keV during the QPO phase all show a negative, soft lag (i.e., akin to reverberation) at all frequencies with a decreasing lag amplitude at low frequencies. Lag-energy spectra at both low frequencies (0.07-0.5 mHz) and the QPO frequency (1.5-2.8 mHz) show similar shapes but with larger amplitudes at lower frequencies (Figure \ref{fig:LFS-LES}). This could arise from a reverberation signal, but the lag amplitude at the lowest frequencies probed is significantly larger than what is commonly measured around $10^6 \, M_\odot$ SMBHs.
    \item We show that strong X-ray jumps began near when the QPO turned on and continues to present day, occurring on the QPO period (Figures \ref{fig:jumps} and \ref{fig:jumps-fvar}). To our knowledge, \src\, is the only AGN to show such strong, repetitive, and rapid jumps in the X-ray band. These jumps are significantly stronger ($\sim 20-70\%$) than the overall QPO amplitude ($\sim 10-20\%$), occur stochastically, and all show the same shape. Current models cannot explain all of the observed properties of the jumps. 
\end{itemize} 

In summary, this work has opened up new windows into the phenomenology of AGN QPOs through detailed spectral-timing analyses of the remarkable mHz-frequency X-ray QPO in \src. Continued monitoring of this source will help to elucidate whether this source is beginning to behave like QPOs in other AGN and BHXRBs, while theoretical efforts to explain the highly coherent extreme X-ray variability will inform our understanding of the inner accretion flow and the dynamics of the X-ray corona. 

\begin{acknowledgments}

MM thanks Norbert Schartel and the entire XMM-Newton team for their approval, scheduling, and execution of many DDT observations of \src. MM thanks Sasha Philippov, Lia Hankla, Chris Reynolds, Andrea Merloni, Chris Done, and Andy Mummery for insightful and thought-provoking discussions regarding the QPO origin. MG is funded by Spanish MICIU/AEI/10.13039/501100011033 and ERDF/EU grant PID2023-147338NB-C21. GM acknowledges support from grant n. PID2023-147338NB-C21 funded by Spanish MICIU/AEI/10.13039/501100011033 and ERDF/EU. CP is supported by INAF Grant 2023 BLOSSOM O.F. 1.05.23.01.13.

\end{acknowledgments}






\facilities{XMM-Newton, NuSTAR}

\software{astropy \citep{AstropyCollaboration2013,AstropyCollaboration2018,AstropyCollaboration2022}, 
\texttt{pyLag} \citep{Wilkins2019},
\texttt{STELA} \citep{Lewin2025}
}

\appendix

\begin{figure*}
    \centering
    \includegraphics[width=\linewidth]{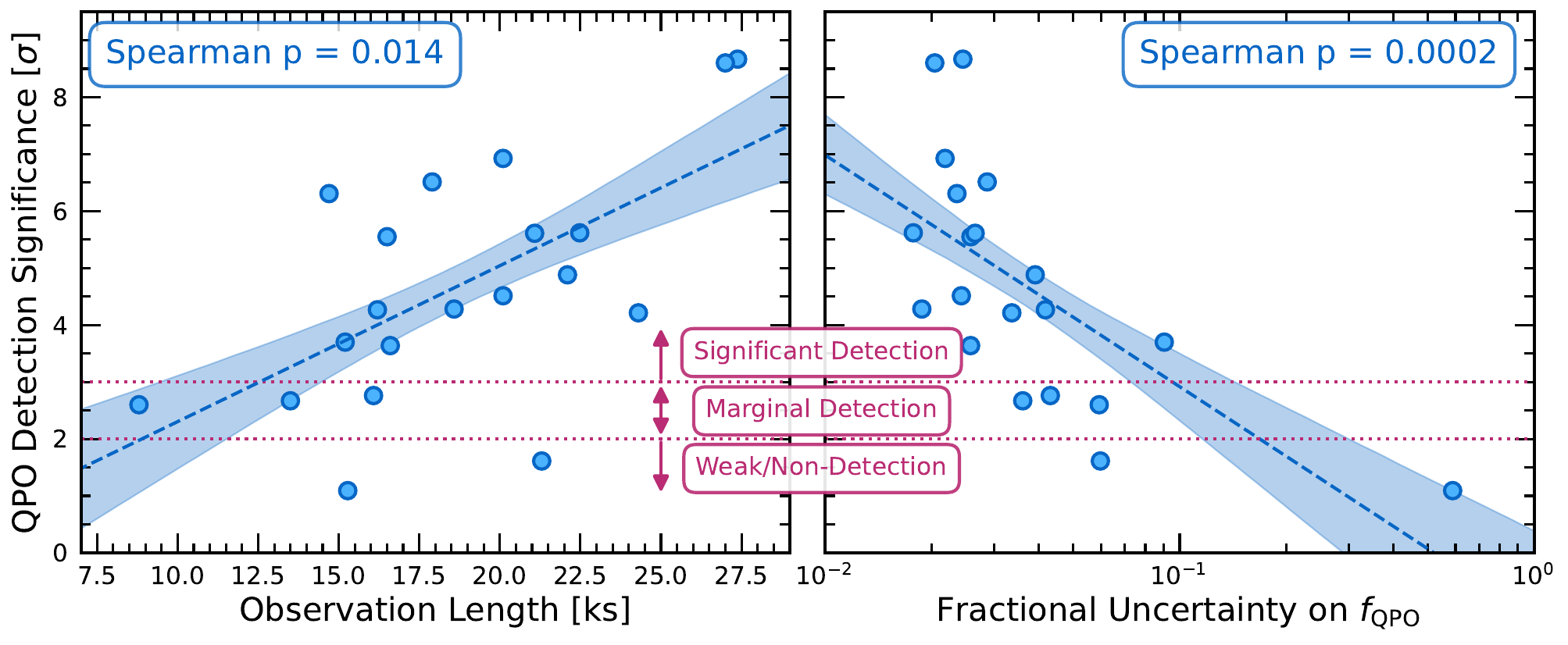}
    \caption{\textit{Left:} Detection significance of the QPO feature as a function of observation length. The trend is statistically significant ($p < 0.05$) and suggests that the detection significance increases with the duration of the observation. There are occasional non-detections where the observational length is expected to be long enough to reach a confident detection, thereby suggesting that the QPO has some level of intermittent variability. \textit{Right:} Detection significance versus fractional uncertainty on the QPO frequency. The trend is statistically significant ($p < 0.05$), suggesting that not reaching a high detection significance decreases our confidence in the observed QPO frequency measurement.}
    \label{fig:sig_trends}
\end{figure*}

In Figure \ref{fig:all_lc_psd} we show all 2-10 keV light curves (left) and their respective PSDs (right). The QPO frequency for each observation, as measured by the best-fit centroid of the Lorentzian, is shown as a pink vertical dashed line in the PSD. Figure \ref{fig:sig_trends} highlights that, for the majority of the observations, the duration of usable data scales with the detection significance of the QPO feature. There are two outliers with relatively long observations ($\gtrsim 15$~ks) that do not show even a marginal QPO detection, thereby suggesting at least some intrinsic variability in the QPO strength over time. Similarly, the fractional uncertainty on the measurement of the QPO frequency scales with the QPO significance.

\begin{figure*}
    \centering
    \includegraphics[width=\linewidth]{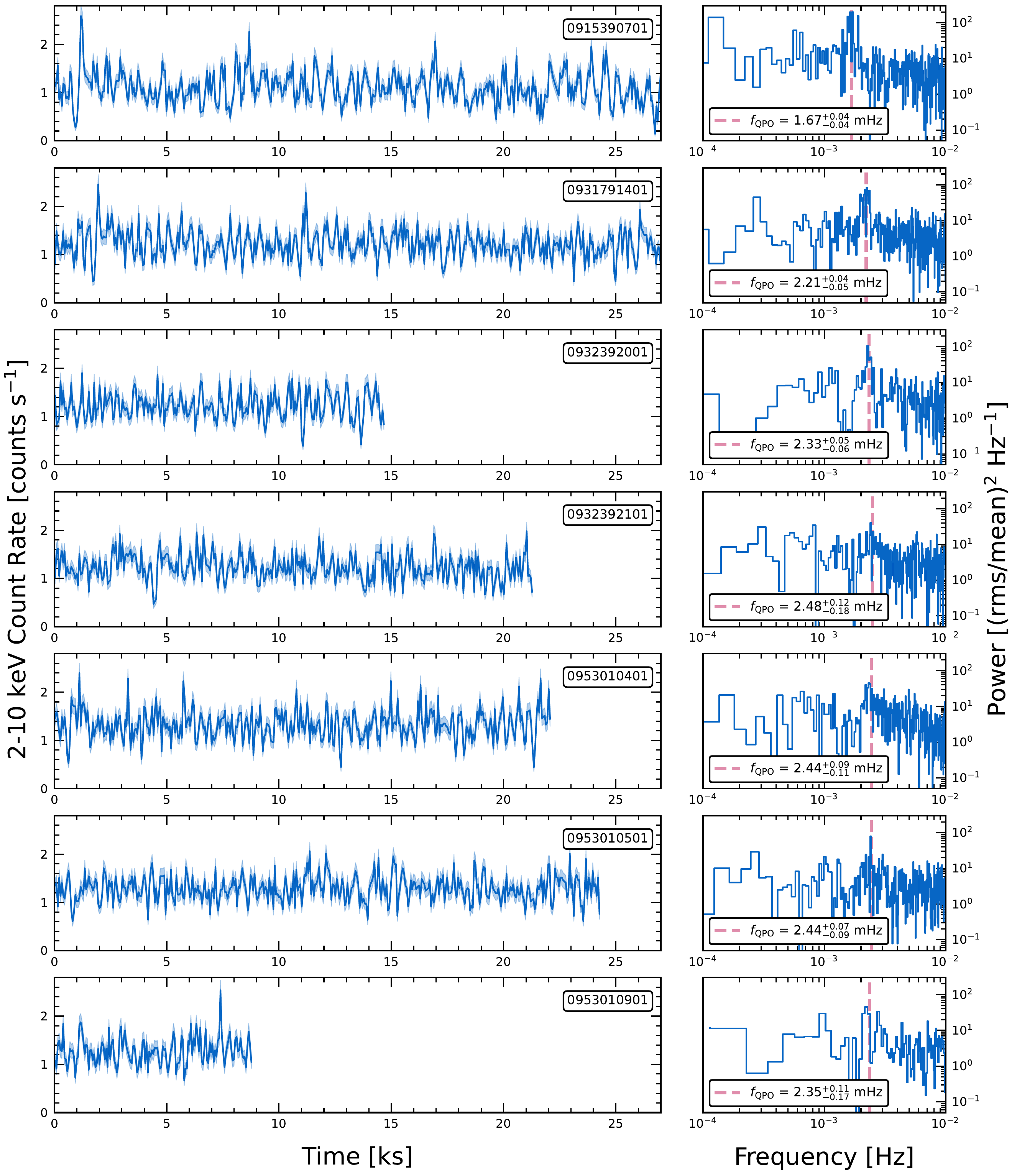}
    \caption{2-10 keV light curves (left) and PSDs (right) for all observations analyzed in this work. For visual clarity, the light curves are binned to 60s bins, with the shaded regions showing the uncertainties. The PSDs are computed with 20s light curves, and the pink vertical dashed line shows the frequency of the QPO, measured from Lorentzian fits. For ease of visual comparison, we fix the $x$- and $y$-axes limits to be the same across all observations. }
    \label{fig:all_lc_psd}
\end{figure*}

\begin{figure*}
    \ContinuedFloat
    \centering
    \includegraphics[width=\linewidth]{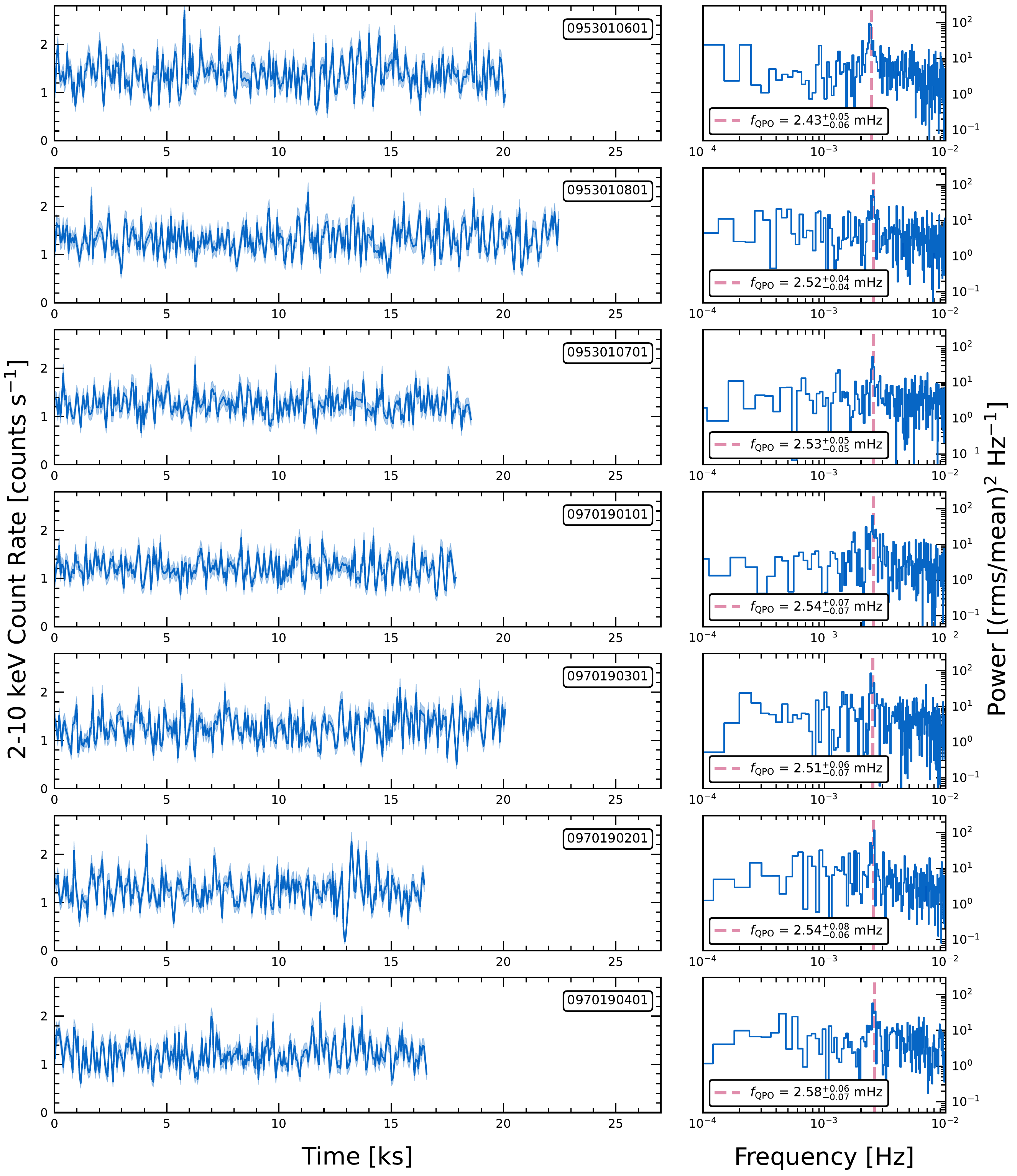}
    \caption{\itshape (continued)}
    \label{fig:all_lc_psd}
\end{figure*}

\begin{figure*}
    \ContinuedFloat
    \centering
    \includegraphics[width=\linewidth]{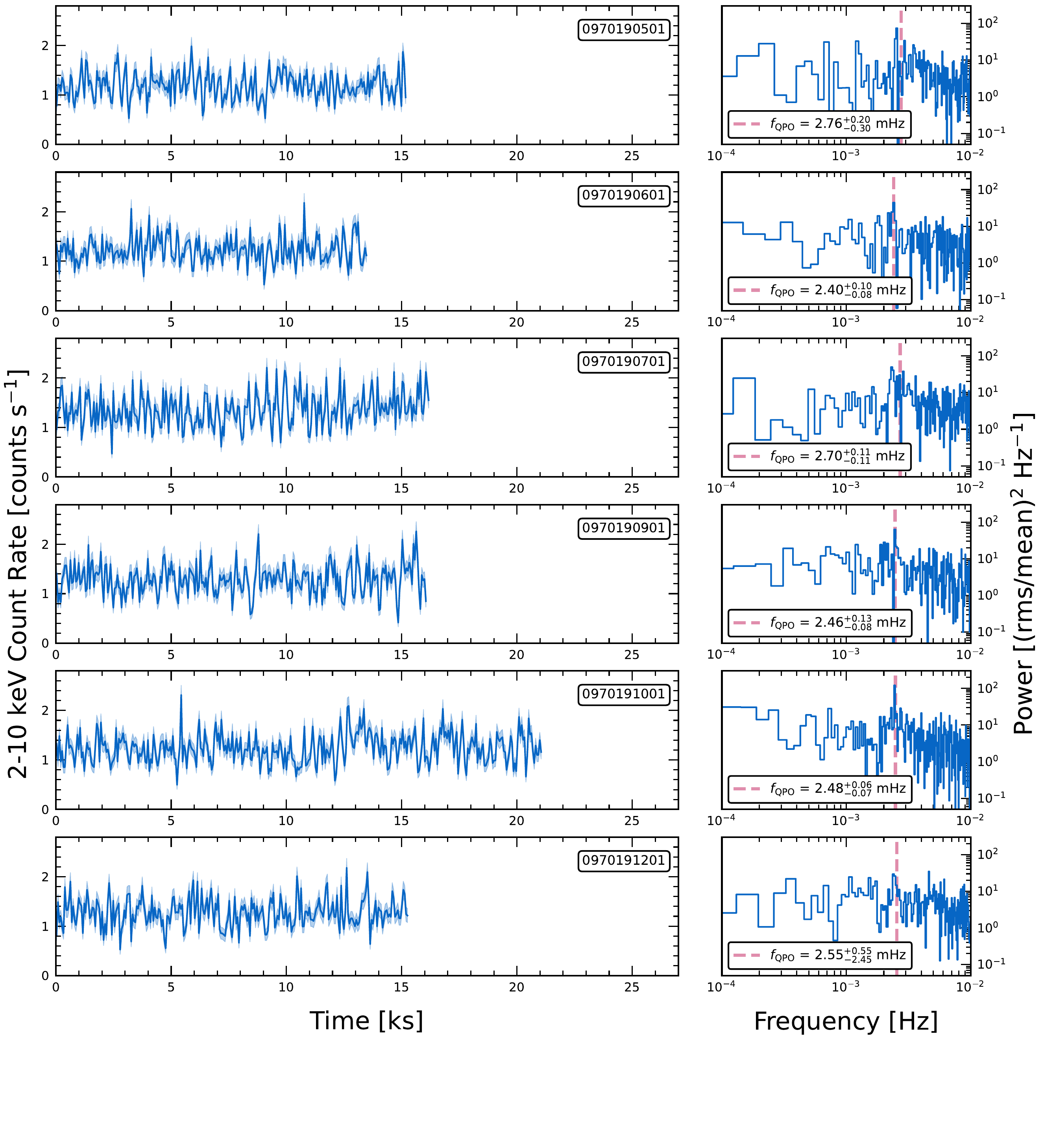}
    \caption{\itshape (continued)}
    \label{fig:all_lc_psd}
\end{figure*}

\bibliography{bib_qpo_followup}{}
\bibliographystyle{aasjournalv7}



\end{document}